\documentclass[hyper,11pt]{JHEP} 

\newcommand{\bea}{\begin{eqnarray}}
\newcommand{\eea}{\end{eqnarray}}
\newcommand{\ba}{\begin{array}}
\newcommand{\ea}{\end{array}}
\newcommand{\bc}{\begin{center}}
\newcommand{\ec}{\end{center}}

\newcommand{\hs}[1]{\hspace{#1 mm}}
\newcommand{\nonum}{\nonumber}

\renewcommand{\a}{\alpha}
\renewcommand{\b}{\beta}
\newcommand{\ga}{\gamma}
\newcommand{\de}{\delta}
\newcommand{\ep}{\epsilon}

\newcommand{\w}{\omega}

\def\pmba{{\pmb{\a}}}
\def\pmbb{{\pmb{\b}}}
\def\pmbga{{\pmb{\ga}}}

\def\pmbde{{\pmb{\de}}}

\newcommand{\Ga}{\Gamma}

\newcommand{\na}{\natural}

\newcommand{\sh}{\sharp}

\newcommand{\half}{\frac{1}{2}}

\def\CB{{\cal B}}

\def\CL{{\cal L}}
\def\CM{{\cal M}}
\def\CG{{\cal G}}

\def\CF{{\cal F}}

\def\CO{{\cal O}}

\def\CC{{\cal C}}

\def\tB{\textsc B}
\def\tC{\textsc C}
\def\tH{\textsc H}
\def\tG{\textsc G}

\def\S{\Sigma}

\def\bfZ{\mathbf Z}

\def\ad{{\dot{\a}}}
\def\bd{{\dot{\b}}}

\def\thetab{{\bar \theta}}

\def\Pih{{\hat \Pi}}

\def\at{{\tilde \a}}
\def\bt{{\tilde \b}}
\def\gat{{\tilde \ga}}
\def\det{{\tilde \de}}

\def\ra{\partial}

\usepackage{amsbsy,amssymb}
\usepackage[dvips]{graphicx}
\usepackage{epsfig}
\newcommand\fverb{\setbox\pippobox=\hbox\bgroup\verb}
\newcommand\fverbdo{\egroup\medskip\noindent%
			\fbox{\unhbox\pippobox}\ }
\newcommand\fverbit{\egroup\item[\fbox{\unhbox\pippobox}]}
\newbox\pippobox
\title{IIB-Branes
and New Spacetime Superalgebras}

\author{Makoto SAKAGUCHI\\
	Yukawa Institute for Theoretical Physics,
        Kyoto University\\
        Sakyo-ku, Kyoto 606-8502, Japan\\
	E-mail: \email{sakaguch@yukawa.kyoto-u.ac.jp}}

\preprint{
          \hepth{9909143}\\
          YITP-99-57\\
}	

\abstract{
We provide a classification of the IIB D$p$- and NS$p$-branes
in which the brane action exists due to a non-trivial class of the
Chevalley-Eilenberg
cohomology of free differential algebras.
We then present a new geometric formulation
of the IIB D$p$- and NS$p$-branes ($p\leq 5$) in which
the manifestly superinvariant Wess-Zumino terms are constructed
in terms of the supersymmetric currents.
The supercurrents are obtained by
using supergroup manifolds corresponding to
the IIB-brane superalgebra,
which is characterized by the
generators of D3-, D5-, NS5- and KK5-branes
in addition to the previously introduced generators of
supertranslations, F- and D-strings.
The charges of D1-, F1- and D3-branes
are related to those of the M-algebra,
but some charges of D5- and NS5-branes are not.
The S-duality of the type-IIB theory
is realized as transformations of the supercurrents
generalizing the SO(2) R-symmetry of the superalgebra.
We thus find that the superalgebra is lifted into twelve-dimensions
with signature (11,1).
}


\begin{document} 

\maketitle 

\section{Introduction}
It is now widely appreciated that super $p$-branes play
an important role in studying non-perturbative superstring physics
\cite{D-branes,p-branes}.
The existence of a $p$-brane is guaranteed by the
existence of the corresponding Wess-Zumino (WZ) term,
which is needed for the $\kappa$-symmetry, or
the matching of
the bosonic and fermionic degrees of freedom
on the worldvolume \cite{AETW}.
The WZ term is determined by a closed $(p+2)$-form
which is shown to be characterized by non-trivial
Chevalley-Eilenberg $(p+2)$-cocycles
on the ordinary superspace \cite{dAT}.
These WZ term was shown to cause topological charges
and modify the super Poincar\'e algebra \cite{dAGIT}.
In the spacetime superalgebra, $p$-branes 
are expressed as $p$-form ``central"
charges \cite{Democracy, Townsend lecture 1997, Hull, Hammer}.

It is known that the WZ action $I_{WZ}=\int\CL_{WZ}$
is superinvariant,
but the integrand $\CL_{WZ}$ is not.
The WZ term $\CL_{WZ}$ is
superinvariant up to a total derivative term.
A manifestly supersymmetric formulation of the Green-Schwarz superstring,
including the WZ term,
was given by Siegel \cite{Siegel} based on a superalgebra \cite{Green}
which is the global limit of a superalgebra \cite{Siegel1}
found by himself.
The superalgebra
is a generalization of super Poincar\'e algebra
and generated by generators of supertranslations $(P_a, Q_\a)$
and those of F-strings $(Z^a, Z^\a)$,
where $Z^\a$ is a new fermionic generator.
Constructing a suitable set of supercurrents on
the corresponding supergroup
manifold, he wrote down the WZ term of
the Green-Schwarz superstring
in a second order expression of the supercurrents.
Using this formulation, $p$-branes ($p>1$),
found in the brane scan \cite{AETW},
were formulated by Bergshoeff and Sezgin \cite{BS}.
Introducing the $p$-brane superalgebras
generated by generators of supertranslation and those of $p$-brane
$(Z^{a_1\cdots a_p},\dots,Z^{\a_1\cdots \a_p})$,
they wrote down the manifestly supersymmetric WZ terms
for the $p$-branes, which are $(p+1)$-th order in the supercurrents
(further discussions on the superalgebra are found in \cite{Alex}).
For the non-minimal type-II theories,
type-II superstrings and IIB D-strings \cite{sakagu1},
IIB $(p,q)$-strings \cite{sakagu2, AHKT}
and IIA D2-brane \cite{CdAIPB} are formulated by introducing
the corresponding superalgebras.
In \cite{sakagu1}, it was pointed out that worldvolume gauge fields
can be expressed in terms of coordinates corresponding to new generators.
This was further investigated in \cite{AHKT, CdAIPB}.

In this paper, we generalize them to incorporate all the
branes in the type-IIB theory.
The type-IIB theory is chiral
with 16+16 Majorana-Weyl supercharges with the same chirality.
The 528 components of the anti-commutator of the supercharges
are distributed as
\bea
{\bf 528=10+10+10+120+126+126+126},
\nonum
\eea
among the components of generators:
translations, D-strings (D9-branes), F-strings
(Neveu-Schwarz (NS) 9-branes),
D3-branes (D7- and NS7-branes), D5-branes,
NS5-branes and Kaluza-Klein (KK) 5-branes,
respectively,
where branes in the parentheses correspond to charges
with a time index.
It is shown that the WZ terms for D$p$- and NS$p$-branes
are characterized by
Chevalley-Eilenberg (CE) cohomology on the ordinary superspace.
The obtained WZ terms are equivalent to those obtained in \cite{APS2},
and the total actions are invariant under the $\kappa$-symmetry.
In order to formulate the WZ action to be manifestly superinvariant,
we introduce new generators with spinorial indices in exchange for
Lorentz indices
of bosonic brane charges of the superalgebra
with the maximal ``central'' extension,
and provide new spacetime superalgebra, the IIB-brane superalgebra.
The IIB-brane superalgebra contains various subalgebras
found before:
N=1 $p$-brane superalgebra \cite{Siegel,BS},
the IIB-superstring superalgebra \cite{sakagu1},
etc.
The D1, F1, D3-brane and KK5-brane charges are obtained by a T-duality
from those in the M-algebra found in \cite{Sezgin},
but the D5- and NS5-brane charges with five spinorial indices are not.
In this sence, the IIB-brane superalgebra is a generalization of the M-algebra.
By using the IIB-brane superalgebra,
we show that the IIB D$p$- and NS$p$-branes ($p\leq 5$)
can be formulated in a manifestly superinvariant way.
Namely, the WZ terms are found to be written down in terms of
the supercurrents on the supergroup manifold
corresponding to the superalgebra.
New coordinates associated with new generators are contained
in a surface term,
by virtue of the fiber bundle structure of our superalgebra
with the ordinary superspace being the base space.
The WZ actions of the superinvariant WZ terms are equivalent to
those obtained in the CE cohomology classification.
Thus, the total actions are $\kappa$-invariant.
The IIB-brane superalgebra enjoys the SO(2) symmetry,
which is a generalization of the  SO(2) R-symmetry,
and thus the S-duality is generalized to include new generators.
The S-duality is shown to be realized geometrically in twelve-dimensions.

This paper is organized as follows.
We begin with showing that the IIB WZ terms
are characterized by the CE cohomology
in sec. 2.
We introduce two free differential algebras,
which turn out to correspond to D-branes and NS-branes.
The obtained WZ terms are shown to be equivalent to those
of the $\kappa$-invariant D/NS $p$-brane actions found before.
In sec. 3, introducing a set of new spacetime superalgebras,
we construct the superinvariant WZ terms for D- and NS-branes
using the supercurrents on the corresponding group manifolds.
The properties of superalgebras and the relations to the
M-algebra are discussed in sec. 4.
It is shown that this IIB-brane superalgebra
enjoys SO(2) symmetry.
The superalgebra is found to be lifted into twelve-dimensions
with signature (11,1),
while realizing the SO(2) symmetry geometrically, in sec. 5.
The last section is devoted to a summary and discussions.

\section{Chevalley-Eilenberg Cohomology Classification of IIB-branes}
The WZ term $\CL_{WZ}$ is known to be
superinvariant up to a total derivative term.
It follows that $(p+2)$-form $h_{p+2}\equiv d\CL_{WZ}$ is
superinvariant.
For the $p$-branes \cite{dAT} and the IIA D$p$-branes \cite{CdAIPB},
the $(p+2)$-form $h_{p+2}$ were characterized by being a non-trivial
CE cohomology $(p+2)$-cocycles on superspace.
In this section, we give a classification of the IIB D$p$-branes
and the IIB NS$p$-branes using the CE cohomology \cite{CE}.

For the type-IIB theory,
the Poincar\'e superalgebra is generated by
the translation generator $P_a$
and 16+16 Majorana-Weyl supercharges $(Q_\a,Q_\at)$
with the same chirality,
as
\footnote{
The notation and conventions are similar to \cite{Democracy}.
The vector indices sare $a=0,1,\dots,9$ and
the spinor indices are $\pmba, \pmbb$.
The charge conjugation matrix is $c_{\a\b}$
and $\ga^{a_1\cdots a_n}=\ga^{[\a_1}\cdots\ga^{a_n]}$. 
}
\bea
\{Q_\pmba,Q_\pmbb\}=(\ga^a1)_{\pmba\pmbb}P_a,\quad
[P_a, Q_\pmba]=[P_a,P_b]=0,
\eea
where $\ga^a$'s are the ten-dimensional $\ga$-matrices satisfying
$\{\ga^a,\ga^b\}=2\eta^{ab}$ and
1 acts on a column vector
$Q_\pmba={Q_\at \choose Q_\a}$.
The left-invariant (LI) Maurer-Cartan one-form $\Pi^A$ is
defined by $\Pi^AT_A=U^{-1}dU$,
where $U$ is a group element and $T^A$ collectively denotes the generators
of the algebra.
Let $(x^a,\theta^\pmba)$ be the coordinates on the corresponding
group manifold,
the LI one-form is expressed as
$
\Pi^a=dx^a+\half(\thetab\ga^a1d\theta)
$ and
$
\Pi^\pmba=d\theta^\pmba
$
and provides a basis for the differential forms on the group manifold.
The LI group vielbein $L_M{}^A$ is read off from $\Pi^A=dZ^ML_M{}^A$,
where $Z^M$ denotes coordinates on the supergroup
manifold.
Similarly, the right-invariant (RI) Maurer-Cartan one-form $\Xi^A$ is
defined by $\Xi^AT_A=dUU^{-1}$,
and the RI group vielbein $R_M{}^A$ is read off from $\Xi^A=dZ^MR_M{}^A$.
The RI generators $Q_A$ of the left-translation,
written as $Q_A=R_A{}^M\ra_M$,
are regarded as the supertranslation generators,
while the LI generators $D_A$ of the right-translation,
written as $D_A=L_A{}^M\ra_M$,
are regarded as the supercovariant derivatives.
The LI one-form $\Pi^A$ is invariant under the supertranslations.

In order to classify the IIB D$p$-branes,
we introduce 
two-form $\CF$, which is the modified field strength
of the NS$\otimes$NS two-form gauge potential
as well as the LI one-form $\Pi^A$.
The non-trivial $(p+2)$-cocycles
are given by closed $(p+2)$-forms
$h_{p+2}(\Pi^a,\Pi^\a,\Pi^\at,\CF)$ which can not be written as the
differential of a $(p+1)$-form constructed from $\Pi^a,\Pi^\a,\Pi^\at$
and $\CF$.
The possible WZ terms will be characterized as
some non-trivial $(p+2)$-cocycles of
the cohomology of a certain free differential algebra (FDA).

Let us consider 
the FDA $\mathfrak{F}$ generated by $\Pi^a,\Pi^\a,\Pi^\at,\CF$
and defined by the structure relations:
\bea
d\Pi^a&=&-\half\Pi^\pmba\Pi^\pmbb(\ga^a1)_{\pmba\pmbb},
\label{FDA1;dPi}\\
d\Pi^\pmba&=&0,
\label{FDA1;dPi alpha}\\
d\CF&=&
\Pi^a\Pi^\pmba\Pi^\pmbb(\ga_a\sigma_3)_{\pmba\pmbb}.
\label{FDA1;dCF}
\eea
The wedge product $\wedge$ is to be understood here and henceforth.
Given a super $q$-form $G$, the exterior derivative acts as follows:
$
d(F\wedge G)=F\wedge dG+(-1)^qdF\wedge G
$.
The nilpotency $dd=0$ is satisfied by the Fierz identity:
$
(\ga_a1)_{(\pmba\pmbb}(\ga^a\sigma_3)_{\pmbga\pmbde)}=0
$.
The superinvariant closed $(p+2)$-form can be expanded
in terms of the two-form $\CF$ as
\bea
h_{p+2}&=&
\sum_{n=0}^{[\frac{p+2}{2}]}\frac{1}{n!}h^{(p+2-2n)}(\Pi^a,\Pi^\pmba)
\CF^n,
\eea
where $h^{(p+2-2n)}$ is a $(p+2-2n)$-form written in terms of
$\Pi^a$ and $\Pi^\pmba$.
Assigning dimension $[\Pi^a]=L$, $[\Pi^\a]=L^{1/2}$, $[\CF]=L^2$,
$[h_{p+2}]=L^{p+1}$,
to avoid introducing extra dimensionful constants,
one writes the $k$-form $h^{(k)}$ with dimension $L^{k-1}$ as
\bea
h^{(k)}&=&a^{(k)}\Pi^{a_1}\cdots\Pi^{a_{k-2}}\Pi^\pmba\Pi^\pmbb
(\ga_{a_1\cdots a_{k-2}}\sigma^{(k)})_{\pmba\pmbb},
\eea
where $a^{(k)}$ is a constant and $\sigma^{(k)}$ is a $2\times 2$ matrix
determined below.
Note that
$\Pi^\pmba\Pi^\pmbb(\ga_{a_1\cdots a_{k-2}}\sigma^{(k)})_{\pmba\pmbb}$
is non-vanishing
when
\bea
\ba{ll}
\sigma^{(k)}=1,\ \sigma_3,\ \sigma_1\ &{\rm for}\ k-2=1\  \bmod\ 4,\\
\sigma^{(k)}=i\sigma_2\ &{\rm for}\ k-2=3\ \bmod\ 4.
\ea
\eea
This immediately implies that there are no non-trivial $(p+2)$-cocycles
when $p={\rm even}$.
This is consistent with the fact that the IIB D and NS $p$-branes
exist only when $p=$odd.
Thus, the closed-ness condition $dh_{p+2}=0$ reduces to conditions
\bea
dh^{(1)}=0,\qquad
dh^{(k)}+h^{(k-2)}d\CF
=0, \quad k=3,5,7,\dots, p+2.
\label{Dp;Bianchi}
\eea
As a result, we find that the non-trivial $h_{p+2}$ is constructed
on the FDA $\mathfrak F$
in terms of
\bea
h^{(1)}&=&
   0,\\
h^{(3)}&=&
   a\Pi^a\Pi^\pmba\Pi^\pmbb(\ga_a\sigma_1)_{\pmba\pmbb},
\label{FDA1;h(3)}\\
h^{(5)}&=&
   a\frac{2}{3!}\Pi^{a_1}\Pi^{a_2}\Pi^{a_3}\Pi^\pmba\Pi^\pmbb
   (\ga_{a_1a_2a_3}i\sigma_2)_{\pmba\pmbb},
\label{FDA1;h(5)}\\
h^{(7)}&=&
   a\frac{2^2}{5!}\Pi^{a_1}\cdots\Pi^{a_5}\Pi^\pmba\Pi^\pmbb
   (\ga_{a_1\cdots a_5}\sigma_1)_{\pmba\pmbb},
\label{FDA1;h(7)}\\
h^{(9)}&=&
   a\frac{2^3}{7!}\Pi^{a_1}\cdots\Pi^{a_7}\Pi^\pmba\Pi^\pmbb
   (\ga_{a_1\cdots a_7}i\sigma_2)_{\pmba\pmbb},
\eea
where $a$ is a normalization constant to be set being 1 below.
In the course to obtain the above results we used Fierz identities.
In $(9+1)$-dimensions,
letting $\phi_i, i=1,2,3,4$, be Grassmann even Majorana-Weyl spinors
with the same chirality,
Fierz identity is expressed as
\bea
(\phi_1\Lambda^1\phi_2)(\phi_3\Lambda^2\phi_4)
&=&\frac{1}{16}\sum_A(\phi_1\CO^A\phi_4)
(\phi_3\Lambda^2\CO^A\Lambda^1\phi_2),
\\
\CO^A&=&\{
\ga^a,
\frac{i}{\sqrt{3!}}\ga^{a_1a_2a_3},
\frac{1}{\sqrt{2\cdot 5!}}\ga^{a_1\cdots a_5}
\}.
\eea
In addition, a useful relation is 
\bea
\ga^{a_1\cdots a_N}=\frac{-(-1)^{\frac{N(N-1)}{2}}}{(10-N)!}
\ep^{a_1\cdots a_Nb_1\cdots b_{10-N}}
\ga_{b_1\cdots b_{10-N}}
\ga^\na,
\eea
where $\ga^\na=\ga^{01\cdots9}$.
In particular, $h_{p+2}$ is closed due to the Fierz identities for
D$p$-branes:
\bea
&\mbox{ D-string}:&(\ga_a)_{(\a\b}(\ga^a)_{\ga)\det}=0,
\label{FDA1;Fierz;D1}\\
&\mbox{ D3-brane}:&
(\ga_c)_{(\a\b}(\ga^{abc})_{\ga)\det}
+2(\ga^{[a})_{(\a\b}(\ga^{b]})_{\ga)\det}=0,
\label{FDA1;Fierz;D3}\\
&\mbox{ D5-brane}:&(\ga_b)_{(\a\b}(\ga^{a_1\cdots a_4b})_{\ga)\det}
+4(\ga^{[a_1})_{(\a\b}(\ga^{a_2a_3a_4]})_{\ga)\det}=0,
\label{FDA1;Fierz;D5}\\
&\mbox{ D7-brane}:&(\ga_b)_{(\a\b}(\ga^{a_1\cdots a_6b})_{\ga)\det}
+6(\ga^{[a_1})_{(\a\b}(\ga^{a_2\cdots a_6]})_{\ga)\det}=0.
\label{FDA1;Fierz;D7}
\eea
The (anti-) symmetrization with a unit weight of indices
in parentheses (square brackets) is understood.
For D9-branes, the Fierz identity is
\bea
&\mbox{ D9-brane}:&(\ga_b)_{(\a\b}(\ga^{a_1\cdots a_8b})_{\ga)\det}
+8(\ga^{[a_1})_{(\a\b}(\ga^{a_2\cdots a_8]})_{\ga)\det}=0
\label{D9}
\eea
while $h^{(11)}$ is formally eleven-dimensional.
$h^{(11)}$ is characterized by
$a^{(11)}=\frac{2^4}{9!}a$ and $\sigma^{(11)}=\sigma_1$.
The Fierz identities (\ref{FDA1;Fierz;D7}) and (\ref{D9})
are obtained from (\ref{FDA1;Fierz;D5}) and (\ref{FDA1;Fierz;D3})
respectively.
One finds that the Fierz identities for D$p$-brane are collectively
expressed as
\bea
(\ga_b)_{(\a\b}(\ga^{a_1\cdots a_{p-1}b})_{\ga)\det}
+(p-1)(\ga^{[a_1})_{(\a\b}(\ga^{a_2\cdots a_{p-1}]})_{\ga)\det}=0.
\eea
In order to see that the obtained $(p+2)$-forms are not CE cohomology trivial,
we note that if there were $(p+1)$-forms $b(\Pi^a,\Pi^\pmba,\CF)$
such that $db=h_{p+2}$, then $b$'s would be Lorentz invariant $(p+1)$-forms
with dimensions $[b]=L^{p+1}$.
One sees that such a $b$ does not exist.

D-brane actions, which are $\kappa$-invariant, have been proposed
in \cite{APS1,APS2,APPS,CGNSW,CGNW,BT}.
We show that our action is $\kappa$-invariant
by relating our action to the $\kappa$-invariant action of \cite{APS2}.
Rescaling supercurrents as
$
\Pi^a\rightarrow k\Pi^a$, $\Pi^\pmba\rightarrow l\Pi^\pmba,
$
where $k=\frac{1}{\sqrt{2}}$ and $kl^2=1$,
(\ref{FDA1;dPi}) is transformed into
$\Pi^a=-\Pi^\pmba\Pi^\pmbb(\ga_a1)_{\pmba\pmbb}$
and (\ref{FDA1;dPi alpha}) and (\ref{FDA1;dCF}) are unchanged.
These are normalizations of supercurrents used in \cite{APS2}.
In this normalization, numerical coefficients of $h^{(k)}, k=3,5,7,9,11$,
are expressed as $1, \frac{1}{3!}, \frac{1}{5!}, \frac{1}{7!}, \frac{1}{9!}$,
respectively.
One find that the $(p+2)$-forms $h_{p+2}$ constructed from these $h^{(k)}$
are identical to those of WZ terms obtained in \cite{APS2}.
It follows that, introducing the Dirac-Born-Infeld action $I_{DBI}$ as
(\ref{DBI action}),
the combined actions, $I=I_{DBI}+I_{WZ}$, are shown to be $\kappa$-invariant
as was done there, but we do not repeat this here.
In this way, we find that $\kappa$-invariant IIB D$p$-brane actions are
characterized by using the CE cohomology.

In order to classify the WZ terms for NS$p$-branes,
we consider the FDA $\mathfrak G$ generated by $\Pi^a,\Pi^\pmba,\CG$
and defined by the structure relations
\bea
d\Pi^a&=&-\half\Pi^\pmba\Pi^\pmbb(\ga^a1)_{\pmba\pmbb},\\
d\Pi^\pmba&=&0,\\
d\CG&=&
-\Pi^a\Pi^\pmba\Pi^\pmbb(\ga_a\sigma_1)_{\pmba\pmbb}
\label{FDA2;dCG}
\eea
where $\CG$ turns out to be the modified field strength of
the Ramond$\otimes$Ramond (R$\otimes$R) two-form
gauge potential.
The closed $(p+2)$-form is expanded as
$
\hat h_{p+2}=
\sum\frac{1}{n!}\hat h^{(p+2-2n)}(\Pi^a,\Pi^\pmba)
\CG^n
$
where $\hat h^{(k)}$ is expressed as
$
\hat h^{(k)}=b^{(k)}\Pi^{a_1}\cdots\Pi^{a_{k-2}}\Pi^\pmba\Pi^\pmbb
(\ga_{a_1\cdots a_{k-2}}\sigma^{(k)})_{\pmba\pmbb}.
$
Solving $d\hat h_{p+2}=0$:
\bea
d\hat h^{(1)}=0,\qquad
d\hat h^{(k)}+\hat h^{(k-2)}d\CG=0,\quad k=2,3,7,\dots,p+2,
\label{NSp;Bianchi}
\eea
we find
\bea
\hat h^{(1)}&=&
   0,\\
\hat h^{(3)}&=&
   b\Pi^a\Pi^\pmba\Pi^\pmbb(\ga_a\sigma_3)_{\pmba\pmbb},
\label{FDA2;h(3)}\\
\hat h^{(5)}&=&
   b\frac{2}{3!}\Pi^{a_1}\cdots\Pi^{a_3}\Pi^\pmba\Pi^\pmbb
   (\ga_{a_1\cdots a_3}i\sigma_2)_{\pmba\pmbb},
\label{FDA2;h(5)}\\
\hat h^{(7)}&=&
   b\frac{2^2}{5!}\Pi^{a_1}\cdots\Pi^{a_5}\Pi^\pmba\Pi^\pmbb
   (\ga_{a_1\cdots a_5}\sigma_3)_{\pmba\pmbb},
\label{FDA2;h(7)}
\\
\hat h^{(9)}&=&
   b\frac{2^3}{7!}\Pi^{a_1}\cdots\Pi^{a_7}\Pi^\pmba\Pi^\pmbb
   (\ga_{a_1\cdots a_7}i\sigma_2)_{\pmba\pmbb},
\eea
where $b$ is a constant to be set being 1 below.
The $(p+2)$-forms $\hat h_{p+2}$ constructed from these $\hat h^{(k)}$'s
are shown to be CE cohomology non-trivial, as before.
As will be seen in the next section,
$\hat h_{p+2}$ is related to $h_{p+2}$ by the S-duality.
We regard the branes associated to the WZ terms characterized by the above
as the IIB NS-branes.
For the 3-brane, we call it D3-brane since the WZ term is transformed to
that of the D3-brane
by a vector-vector duality on the worldvolume gauge field (see sec. 3.2).
The 7-branes will be regarded as the NS7-branes.

$\hat h_{p+2}$ is closed due to Fierz identities :
\bea
&\mbox{ F-string}: &(\ga_a)_{(\a\b}(\ga^a)_{\ga\de)}=0,
\label{FDA2;Fierz;F1}
\\
&\mbox{ D3-brane}: &(\ga_c)_{(\a\b}(\ga^{abc})_{\ga)\det}
   +2(\ga^{[a})_{(\a\b}(\ga^{b]})_{\ga)\det}=0, \\
&\mbox{ NS5-brane}: &
  \left\{\ba{l}
   (\ga_b)_{(\a\b}(\ga^{a_1\cdots a_4b})_{\ga\de)}=0,\\
   (\ga_b)_{\a\b}(\ga^{a_1\cdots a_4b})_{\gat\det}
   -(\ga_b)_{\gat\det}(\ga^{a_1\cdots a_4b})_{\a\b}
   +16(\ga^{[a_1})_{\a\gat}(\ga^{a_2a_3a_4]})_{\b\det}=0,
  \ea\right.
\label{FDA2;Fierz;NS5} \\
&\mbox{ NS7-brane}: &(\ga_b)_{(\a\b}(\ga^{a_1\cdots a_6b})_{\ga)\det}
   +6(\ga^{[a_1\cdots a_5})_{(\a\b}(\ga^{a_6]})_{\ga)\det}=0.
\label{FDA2;Fierz;NS7}
\eea
In the second line of (\ref{FDA2;Fierz;NS5}),
the symmetrization of $(\a,\b)$ and that of $(\gat,\det)$
are understood.
Note that (\ref{FDA1;Fierz;D7}) is different from (\ref{FDA2;Fierz;NS7})
while $h^{(9)}$ is the same as $\hat h^{(9)}$.
The identity (\ref{FDA2;Fierz;NS7}) is obtained from
(\ref{FDA1;Fierz;D5}) and the first line of (\ref{FDA2;Fierz;NS5}).

For NS9-branes, the Fierz identity is
\bea
&\mbox{ NS9-brane}:&
\left\{
\ba{l}
(\ga_b)_{(\a\b}(\ga^{a_1\cdots a_8b})_{\ga\de)}=0,\\
(\ga_b)_{\a\b}(\ga^{a_1\cdots a_8b})_{\gat\det}
-(\ga_b)_{\gat\det}(\ga^{a_1\cdots a_8b})_{\a\b}
+32(\ga^{[a_1})_{\a\gat}(\ga^{a_2\cdots a_8]})_{\b\det}=0,
\ea\right.
\label{NS9}
\eea
while $h^{(11)}$ is formally eleven-dimensional.
One finds that $h^{(11)}$ is characterized by
$b^{(11)}=\frac{2^4}{9!}b$ and $\sigma^{(11)}=\sigma_3$.
The second line of (\ref{NS9}) is obtained from
\bea
(\ga^{[a})_{\a\b}(\ga^{b]})_{\gat\det}
+(\ga^c)_{\a\gat}(\ga^{ab}{}_{c})_{\b\det}=0,
\eea
where the symmetrization of $(\a,\b)$ and that of $(\gat,\det)$
are understood.
We find that the Fierz identities for NS/D $p$-branes
are collectively expressed as
\bea
(\ga^{a_1\cdots a_{p-1}b}i\sigma_2\tau^{\frac{p+1}{2}})_{(\pmba\pmbb}
(\ga^b1)_{\pmbga\pmbde)}
+(p-1)(\ga^{[a_1\cdots a_{p-2}}i\sigma_2\tau^{\frac{p-1}{2}})_{(\pmba\pmbb}
(\ga^{a_{p-1}]}\tau)_{\pmbga\pmbde)}=0,
\eea
where $\tau=\sigma_1$ for NS$p$-branes and $\tau=\sigma_3$ for D$p$-branes.
In this way, we have classified the WZ terms for NS$p$-branes.
Denoting the obtained WZ action as $\hat I_{WZ}$
and introducing $\hat I_{DBI}$ as (\ref{DBI;NS}),
one expects that the combined actions, $\hat I=\hat I_{DBI}+\hat I_{WZ}$,
are shown to be $\kappa$-invariant.
In fact, $\hat I$
can be rewritten as the $\kappa$-invariant actions $I$
by the redefinition of supercurrents: $\Pi^\pmba\rightarrow\sigma\Pi^\pmba$,
where $\sigma=\frac{\pm 1}{\sqrt{2}}(1+i\sigma_2)$,
as will be seen in sec.3.
We expect that the same procedure employed in showing the $\kappa$-invariance
of D$p$-branes will work here.

We have classified closed superinvariant $(p+2)$-forms $h_{p+2}$
and $\hat h_{p+2}$,
and then the WZ term $\CL_{WZ}$ for D$p$-branes
and $\hat\CL_{WZ}$ for NS$p$-branes, respectively.
In the next section, we construct the WZ terms
for D$p$- and NS$p$-branes ($p\leq 5$)
in a manifestly supersymmetric way.
\section{Supersymmetric Wess-Zumino Terms}

We construct the manifestly superinvariant WZ terms
using the supercurrents on the supergroup manifolds
of the corresponding superalgebras.
We discuss the relations of these superalgebras
to the M-algebra and the IIA superalgebra in the next section.

The D$p$-brane action is composed of the Dirac-Born-Infeld (DBI) action
and the WZ action.
The  DBI action is (background scalars are omitted)
\bea
I_{DBI}=\int_{\CM_{p+1}} d^{p+1}\xi
\sqrt{-{\rm det}(\Pi^a_i\Pi^b_j\eta_{ab}-\CF_{ij})}.
\label{DBI action}
\eea
where $\CM_{p+1}$ is the worldvolume of the D$p$-brane
and
$\Pi^a_i$ is the pullback to the worldvolume of
the supercurrent corresponding to translations.
The two-form $\CF$ is the modified field strength of the NS$\otimes$NS
two-form gauge potential
$\tB^{(2)}$, and defined by $\CF=db-\tB^{(2)}$,
where $b$ is the Born-Infeld U(1) gauge field on the worldvolume.
The DBI action is manifestly supersymmetric.

The WZ action for D$p$-branes is defined as
\bea
I_{WZ}=\int_{\CM_{p+1}} e^\CF\wedge \CC,
\qquad
\CC&=&\bigoplus_{n}\CC^{(n)},
\eea
where $\CC^{(n)}$ is characterized by the pullback to the worldvolume of
the $n$-form R$\otimes$R gauge potential $\tC^{(n)}$.\footnote{
We do not distinguish spacetime gauge potentials and the pullback to
the worldvolume of them,
but use the same characters, throughout this paper.
}
The WZ action $I_{WZ}$ is superinvariant.
The invariance is however not manifest,
since the integrand defined by $\CL_{WZ}\equiv[e^\CF\wedge \CC]_{p+1}$
is not superinvariant.
The WZ term is said to be {\it quasi}-superinvariant;
superinvariant {\it up to a total divergence},
$\de \CL_{WZ}=d\CO$.
We show that the WZ term for D$p$-branes can be constructed to be manifestly
superinvariant
in terms of supersymmetric currents on the corresponding supergroups.
To do this, we recall that
we have introduced $h_{p+2}$ as $h_{p+2}=d\CL_{WZ}$ in the previous section.
It follows that
\bea
h^{(3)}&=&d\CC^{(2)},\\
h^{(5)}&=&d\CC^{(4)}+d\CF\CC^{(2)},\\
h^{(7)}&=&d\CC^{(6)}+d\CF\CC^{(4)},\\
h^{(9)}&=&d\CC^{(8)}+d\CF\CC^{(6)}.
\eea
We determine
$\CC^{(k)}, k=2,4,6,$
by solving the above equations recursively
in terms of supersymmetric currents on the supergroups manifolds
corresponding to spacetime superalgebras for D-strings, D3-branes
and D5-branes, respectively.

For NS$p$-branes,
the DBI action is expressed as (omitting the background scalar fields)
\bea
\hat I_{DBI}=\int_{\CM_{p+1}} d^{p+1}\xi
\sqrt{-{\rm det}(\Pi^a_i\Pi^b_j\eta_{ab}-\CG_{ij})},
\label{DBI;NS}
\eea
and is manifestly superinvariant.
The $\CG$ is defined by $\CG=dc-\tC^{(2)}$,
where $\tC^{(2)}$ is the R$\otimes$R two-form gauge potential
and $c$ is a worldvolume gauge field.
Instead of the DBI action, one can use the kinetic term of
the Green-Schwarz action for F-strings,
which is manifestly supersymmetric too.
We define WZ action as
\bea
\hat I_{WZ} &=&\int_{\CM_{p+1}}e^\CG\wedge \CB,\qquad
\CB=
 \CB^{(0)}- \CB^{(2)}+ \CB^{(4)}
- \CB^{(6)}+ \CB^{(8)}- \CB^{(10)}.
\eea
We have introduced $\hat h_{p+2}$ as $\hat h_{p+2}=d\hat\CL_{WZ}$
in the previous section.
It follows that
\bea
\hat h^{(3)}&=&-d\CB^{(2)},\\
\hat h^{(5)}&=&d\CB^{(4)}-d\CG\CB^{(2)},\\
\hat h^{(7)}&=&-d\CB^{(6)}+d\CG\CB^{(4)},\\
\hat h^{(9)}&=&d\CB^{(8)}-d\CG\CB^{(6)}.
\eea
We show that
$\CB^{(k)}, k=2,4,6$, can be constructed
in terms of supersymmetric currents on the supergroups manifolds
corresponding to spacetime superalgebras for F-strings, D3-branes and
NS5-branes, respectively.
\subsection{D- and F-strings}
In order for the present paper to be self-contained,
we begin with describing the IIB D- and F-strings following
\cite{sakagu1,sakagu2}.

For D-strings, we introduce LI Maurer-Cartan (MC) equations
corresponding to supertranslations:
\bea\ba{rcl}
d\Pi^a&=&-\half\Pi^\pmba\Pi^\pmbb(\ga^a1)_{\pmba\pmbb},\\
d\Pi^\pmba&=&0,
\ea
\label{alg;supertranslation}
\eea
and D-strings:
\bea
\ba{rcl}
d{\Pi_a}'&=&-\half\Pi^\pmba\Pi^\pmbb(\ga_a\sigma_1)_{\pmba\pmbb},\\
d{\Pi_\pmba}'&=&-\Pi^\pmbb{\Pi_a}'(\ga^a1)_{\pmba\pmbb}
-\Pi^\pmbb\Pi^a(\ga_a\sigma_1)_{\pmba\pmbb},
\ea
\label{alg;D1}
\eea
which are T-dual to the MC equations for D0- and D2-branes
obtained from the M-algebra as will be seen in sec. 4.2.
MC equations
contain equivalent information about the algebra:
\bea
[D_A,D_B\}={f_{AB}}^CD_C
\Longleftrightarrow
d\Pi^C=-\half\Pi^B\wedge\Pi^A{f_{AB}}^C,
\eea
where $D_A$ is dual to $\Pi^A$ and denotes
the LI generators of the right-transformation
collectively.
The structure constants of the supersymmetry algebra generated by
the RI generators $Q_A$ of the left-transformation
is identical to those of the algebra generated by
the LI generators $D_A$ of the right-transformation
up to a sign.
Keeping this in mind, we call the MC equations of the LI one-forms
the dual to the superalgebra for short, throughout this paper.
The Jacobi identities of the algebra
are satisfied if and only if the integrability conditions
$d^2=0$ on the dual forms hold.
We call the superalgebra corresponding to
(\ref{alg;supertranslation}) and (\ref{alg;D1})
the D-string superalgebra.
The supersymmetric WZ term $\CL_{WZ}=\CC^{(2)}$ for D-strings
is constructed in terms of these supercurrents
on the corresponding supergroup manifold as\footnote{
This (pullback of the) two-form gauge potential $\CC^{(2)}$
is not used for the D-string action,
rather is used for $(p,q)$-string action.
We do not distinguish between them throughout this paper. 
}
\bea
\CC^{(2)}&=&
-\Pi^a{\Pi_a}'-\half\Pi^\pmba{\Pi_\pmba}'.
\label{C2}
\eea
In fact, one can see that
$
d\CC^{(2)}=\Pi^a\Pi^\pmba\Pi^\pmbb(\ga_a\sigma_1)_{\pmba\pmbb},
$
and then $h_3=h^{(3)}$, which is (\ref{FDA1;h(3)}) given by
the FDA $\mathfrak F$,
is obtained.
This is consistent with (\ref{FDA2;dCG})
of the FDA $\mathfrak G$.

In turn for F-strings,
MC equations for supertranslation (\ref{alg;supertranslation})
and F-strings:
\bea
\ba{rcl}
d\Pi_a&=&+\half\Pi^\pmba\Pi^\pmbb(\ga_a\sigma_3)_{\pmba\pmbb}
\\
d\Pi_\pmba&=&+\Pi^\pmbb\Pi^a(\ga_a\sigma_3)_{\pmba\pmbb}
-\Pi^\pmbb\Pi_a(\ga^a1)_{\pmba\pmbb},
\ea
\label{alg;F1}
\eea
are needed.
We call the corresponding superalgebra the F-string superalgebra.
In terms of these supercurrents on the corresponding supergroup manifold,
supersymmetric WZ term $\hat \CL_{WZ}=-\CB^{(2)}$ for F-strings
is constructed by
\bea
\CB^{(2)}&=&
-\Pi^a{\Pi_a}-\half\Pi^\pmba{\Pi_\pmba}.
\label{B2}
\eea
In fact, one can get
$
d \CB^{(2)}=-\Pi^a\Pi^\pmba\Pi^\pmbb(\ga_a\sigma_3)_{\pmba\pmbb},
$
and then $\hat h_3=\hat h^{(3)}$, which is (\ref{FDA2;h(3)})
obtained from the FDA $\mathfrak G$.
This is consistent with (\ref{FDA1;dCF}) of the FDA
$\mathfrak F$.

New coordinates for D- and F-string charges
are contained in a total derivative term.
$h^{(3)}$ and $\hat h^{(3)}$ can be regarded as
the pullback to the worldvolume of the field strengths of the spacetime
gauge potentials,
$\tH^{(3)}=d\tC^{(2)}$ and $-\tG^{(3)}=-d\tB^{(2)}$,
which are invariant under the spacetime gauge transformations
\bea
\de\tC^{(2)}=d\Lambda^{(1)},\qquad
\de\tB^{(2)}=d\Omega^{(1)}.
\label{gauge tfn;1}
\eea
We regard $\CB^{(2)}$ and $\CC^{(2)}$ as (the pullback to the worldvolume of)
the spacetime gauge potentials $\tB^{(2)}$ and $\tC^{(2)}$, respectively.
The S-duality
\bea
\tB^{(2)}\rightarrow -\tC^{(2)},\qquad
\tC^{(2)}\rightarrow \tB^{(2)},
\label{S-duality;1}
\eea
maps the F-string WZ term $\hat \CL_{WZ}$ to the D-string one $\CL_{WZ}$.
The S-duality transformations (\ref{S-duality;1})
can be expressed as a set of transformations of the supercurrents:
\bea
\Pi^a\rightarrow\Pi^a,\quad
\Pi_a\rightarrow -{\Pi_a}',\quad
{\Pi_a}'\rightarrow {\Pi_a},\quad
{\Pi^\pmba}\rightarrow \sigma{\Pi^\pmba},\quad
{\Pi_\pmba}\rightarrow -\sigma{\Pi_\pmba}',\quad
{\Pi_\pmba}'\rightarrow \sigma{\Pi_\pmba},
\label{S-duality;currents;1}
\eea
where $\sigma=\frac{\pm 1}{\sqrt{2}}(1+i\sigma_2)$.
These transformations are regarded as the generalized R-symmetry
which acts not only on $\Pi^\pmba$ but also on $\Pi_\pmba{}'$ and $\Pi_\pmba$,
as will be seen in the sec. 4.2.

\subsection{D3-branes}\label{WZ;D3}

For D3-branes, we introduce MC equations:
\bea
\ba{rcl}
d\Pi_{abc}&=&
   \half\Pi^\pmba\Pi^\pmbb(\ga_{abc}i\sigma_2)_{\pmba\pmbb}
,\\
d\Pi_{ab\pmba}&=&
   \Pi^\pmbb\Pi_{abc}(\ga^c1)_{\pmba\pmbb}
   -\Pi^\pmbb\Pi^c(\ga_{abc}i\sigma_2)_{\pmba\pmbb}
   +2\Pi^\pmbb\Pi_a(\ga_b\sigma_1)_{\pmba\pmbb}
   +2\Pi^\pmbb{\Pi_a}'(\ga_b\sigma_3)_{\pmba\pmbb}
,\\
d\Pi_{a\pmba\pmbb}&=&
   -\half\Pi_{abc}\Pi^b(\ga^c1)_{\pmba\pmbb}
   +\frac{1}{4}\Pi_{ab\pmbga}\Pi^\pmbga(\ga^b1)_{\pmba\pmbb}
   +2\Pi_{ab\pmba}\Pi^\pmbga(\ga^b1)_{\pmbb\pmbga}
   -\half\Pi^b\Pi^c(\ga_{abc}i\sigma_2)_{\pmba\pmbb}
\\&&
   -\half\Pi^b{\Pi_b}'(\ga_a\sigma_3)_{\pmba\pmbb}
   -\half\Pi^b\Pi_b(\ga_a\sigma_1)_{\pmba\pmbb}
   +\Pi^b{\Pi_a}'(\ga_b\sigma_3)_{\pmba\pmbb}
   +\Pi^b\Pi_a(\ga_b\sigma_1)_{\pmba\pmbb}
\\&&
   -\frac{1}{4}\Pi^\pmbga{\Pi_\pmbga}'(\ga_a\sigma_3)_{\pmba\pmbb}
   -\frac{1}{4}\Pi^\pmbga\Pi_\pmbga(\ga_a\sigma_1)_{\pmba\pmbb}
   -\half\Pi_b{\Pi_a}'(\ga^b1)_{\pmba\pmbb}
   -\half\Pi_a{\Pi_b}'(\ga^b1)_{\pmba\pmbb}
\\&&
   -2\Pi^\pmbga\Pi_\pmba(\ga_a\sigma_1)_{\pmbb\pmbga}
   -2\Pi^\pmbga{\Pi_\pmba}'(\ga_a\sigma_3)_{\pmbb\pmbga}
,\\
d\Pi_{\pmba_1\pmba_2\pmba_3}&=&
   2\Pi^a\Pi_{ab\pmba_1}(\ga^b1)_{\pmba_2\pmba_3}
   +\Pi^\pmbb\Pi_{a\pmbb\pmba_1}(\ga^a1)_{\pmba_2\pmba_3}
   +5\Pi^\pmbb\Pi_{a\pmba_1\pmba_2}(\ga^a1)_{\pmba_3\pmbb}
   -3{\Pi_a}'\Pi_{\pmba_1}(\ga^a1)_{\pmba_2\pmba_3}
\\&&
   +3\Pi_a{\Pi_{\pmba_1}}'(\ga^a1)_{\pmba_2\pmba_3}
   -5\Pi^a{\Pi_{\pmba_1}}'(\ga_a\sigma_3)_{\pmba_2\pmba_3}
   -5\Pi^a\Pi_{\pmba_1}(\ga_a\sigma_1)_{\pmba_2\pmba_3}
,\ea
\label{alg;D3}
\eea
in addition to MC equations for supertranslations
(\ref{alg;supertranslation}),
F-strings (\ref{alg;F1}) and D-strings (\ref{alg;D1}).
We call the superalgebra corresponding to
(\ref{alg;supertranslation}),
(\ref{alg;F1}),
(\ref{alg;D1}) and (\ref{alg;D3}) the D3-brane superalgebra.
The MC equations (\ref{alg;D3}) are T-dual to those of D2-branes
and D4-branes obtained from the M-algebra as will be seen in sec. 4.3.

We find that the supersymmetric WZ term is constructed as
\bea
\CL_{WZ}=\CC^{(4)}+\CF\wedge \CC^{(2)},
\eea
where
\bea
\CC^{(4)}&=&
   \tC^{(4)}
   +\half\tC^{(2)}\tB^{(2)},
\label{C(4)}\\
\tC^{(4)}&=&
   \frac{1}{6}\Pi^a\Pi^b\Pi^c\Pi_{abc}
   -\frac{29}{140}\Pi^a\Pi^b\Pi^\pmba\Pi_{ab\pmba}
   -\frac{3}{35}\Pi^a\Pi^\pmba\Pi^\pmbb\Pi_{a\pmba\pmbb}
   +\frac{1}{140}\Pi^\pmba\Pi^\pmbb\Pi^\pmbga\Pi_{\pmba\pmbb\pmbga}
\nonum\\&&
   -\frac{3}{35}\Pi^a\Pi_a \tC^{(2)}
   +\frac{3}{35}{\Pi^a}{\Pi_a}' \tB^{(2)},
\eea
and $\tC^{(2)}$ and $\tB^{(2)}$ are found in (\ref{C2}) and (\ref{B2}),
respectively.
In fact, this implies
\bea
d \CC^{(4)}-d \CB^{(2)} \CC^{(2)}=
\frac{1}{3}\Pi^{a_1}\Pi^{a_2}\Pi^{a_3}\Pi^\pmba\Pi^\pmbb
(\ga_{a_1a_2a_3}i\sigma_2)_{\pmba\pmbb},
\eea
and the right hand side is the $h^{(5)}$ found in (\ref{FDA1;h(5)}),
that is $d\CL_{WZ}=h_5$.
This supersymmetric WZ term is equivalent to the
WZ term obtained in sec. 2 up to a total derivative.
New coordinates are contained in a surface term as was shown
in the case of F- and D-strings.
To see this, we introduce coordinates $x^a$ and $\theta^\pmba$
on the supergroup manifold
associated with the generators for the supertranslations $P_a$ and $Q_\pmba$,
respectively.
One finds that for any parametrization of the supergroup manifold
the supercurrents $\Pi^a$ and $\Pi^\pmba$ are expressed as
$\Pi^a=dx^a+\half(\thetab\ga^a1d\theta)$ and $\Pi^\pmba=d\theta^\pmba$,
because of
the fiber bundle structure of our superalgebra
with the ordinary superspace
characterized by (\ref{alg;supertranslation})
being the base space.
This implies that the terms containing new coordinates in $\CL_{WZ}$
are eliminated by an exterior derivative $d$
and thus contained in a surface term.
This is a universal property of our formulation.

We can construct the WZ term for D3-branes in a different form.
The supersymmetric WZ term is
\bea
\hat \CL_{WZ}= \CB^{(4)}-\CG\wedge  \CB^{(2)}
\eea
where
\bea
 \CB^{(4)}&=&\tC^{(4)}
   -\half \tC^{(2)} \tB^{(2)}.
\label{B(4)}
\eea
In fact, one can show that
\bea
d \CB^{(4)}+d \CC^{(2)} \CB^{(2)}=
\frac{1}{3}\Pi^{a_1}\Pi^{a_2}\Pi^{a_3}\Pi^\pmba\Pi^\pmbb
(\ga_{a_1a_2a_3}i\sigma_2)_{\pmba\pmbb},
\eea
and the right hand side is the $\hat h^{(5)}$ found in (\ref{FDA2;h(5)}).
We have shown that $d\hat\CL_{WZ}=h_5$
and 
new coordinates are contained in a surface term as was
seen above.
The superinvariant WZ terms $\CL_{WZ}$ and $\hat \CL_{WZ}$
are equivalent to those obtained in sec. 2 up to a total derivative,
and thus the total actions are $\kappa$-invariant.

Now let us comment on the relation between $\CL_{WZ}$ and $\hat\CL_{WZ}$.
First, $I=I_{DBI}+I_{WZ}$ is transformed to $\hat I=\hat I_{DBI}+\hat I_{WZ}$
by a vector-vector duality transformation \cite{Tseytlin,GG},
$b\leftrightarrow c$.
We add a Lagrange multiplier term
$\half \Lambda^{ij}(\CF_{ij}-2\ra_ib_j+\tB^{(2)}_{ij})$ to $I$.
Solving for $b_i$ implies that $\Lambda^{ij}=-\epsilon^{ijkl}\ra_kc_l$,
where $c_i$ is a vector field dual to $b_i$
and turns out to be the worldvolume gauge field associated with the
R$\otimes$R two-form $\tC^{(2)}$.
One obtains $\hat I$ after
eliminating $\CF_{ij}$ using the equations of motion.
The term $\half\tC^{(2)}\tB^{(2)}$
emerged in (\ref{C(4)}) and (\ref{B(4)}) in our formulation.
This is consistent with the spacetime gauge transformations
as will be seen below.

Second, the action $\hat I$ is mapped to the action $I$
under the S-duality transformations:
(\ref{S-duality;1}) and
\bea
&
\tC^{(4)}\rightarrow\tC^{(4)},
&
\label{S-duality;3}\\
&
\CG\rightarrow\CF,\quad\CF\rightarrow-\CG,\quad
c\rightarrow b,\quad b\rightarrow -c.
&
\label{S-duality;wv}
\eea
These two imply that $I$ is self-dual by itself
under the S-duality and the worldvolume vector-vector duality transformation.
The same is true for $\hat I$.

Third, $h^{(5)}$ and $\hat h^{(5)}$ are identical
from the expression in terms of the supercurrents
as well as the spacetime gauge potentials.
These are regarded as the pullback to the
worldvolume of the field strength $\tH^{(5)}$
of the spacetime gauge potential
$ \tC^{(4)}$
\bea
\tH^{(5)}=
  d \tC^{(4)}
  -\half\tC^{(2)}d\tB^{(2)}
  +\half\tB^{(2)}d\tC^{(2)}.
\eea
The Bianchi identity is $d\tH^{(5)}-\tH^{(3)}\tG^{(3)}=0$,
which corresponds to $dh^{(5)}+h^{(3)}d\CF=0$ of (\ref{Dp;Bianchi})
or $d\hat h^{(5)}+\hat h^{(3)}d\CG=0$ of (\ref{NSp;Bianchi}).
$\tH^{(5)}$ is invariant under the spacetime gauge transformations:
(\ref{gauge tfn;1}) and
\bea
\de\tC^{(4)}=d\Lambda^{(3)}
  +\half d\Lambda^{(1)}\tB^{(2)}
  -\half d\Omega^{(1)}\tC^{(2)}.
\label{gauge tfn;2}
\eea
$\CF$ and $\CG$ are gauge invariant by definition
of the gauge transformations
of the worldvolume gauge fields $b$ and $c$: $\de b=d\w +\Omega^{(1)}$
and $\de c=d\lambda +\Lambda^{(1)}$, where $\w$ and $\lambda$ are
worldvolume scalars.
The gauge invariances of $I_{WZ}$ and $\hat I_{WZ}$
are equivalent to the gauge quasi-invariances of $\CL_{WZ}$
and $\hat\CL_{WZ}$,
and then to the gauge invariances, $\de h^{(5)}=\de h^{(3)}=0$
and $\de \hat h^{(5)}=\de \hat h^{(3)}=0$.
These are guaranteed by the gauge invariances of $\tH^{(5)}$, $\tH^{(3)}$
and
$\tG^{(3)}$.

Finally, the S-duality transformations (\ref{S-duality;3})
can be expressed as a set of transformations of supercurrents:
(\ref{S-duality;currents;1}) and
\bea
\Pi_{abc}\rightarrow \Pi_{abc},\quad
\Pi_{ab\pmba}\rightarrow\Pi_{ab(\sigma\pmba)},\quad
\Pi_{a\pmba\pmbb}\rightarrow\Pi_{a(\sigma\pmba)(\sigma\pmbb)},\quad
\Pi_{\pmba\pmbb\pmbga}\rightarrow
    \Pi_{(\sigma\pmba)(\sigma\pmbb)(\sigma\pmbga)},
\label{S-duality;currents;3}
\eea
where $\Pi_{ab(\sigma\pmba)}=
\sigma {\Pi_{ab\at} \choose \Pi_{ab\a} }$ etc.
These transformations are shown to be the generalized R-symmetry,
and thus the S-duality is realized as an automorphism of the D3-brane
superalgebra, as will be seen in the sec. 4.3.

\subsection{D5- and NS5-branes}
For D5-branes, we begin with MC equations
listed in the appendix A,
in addition to those for supertranslations (\ref{alg;supertranslation}),
F-strings (\ref{alg;F1}), D-strings (\ref{alg;D1}) and D3-branes
(\ref{alg;D3}).
We call the corresponding superalgebra the D5-brane superalgebra.
We find that the supersymmetric WZ terms for D5-branes can be constructed
in terms of the supercurrents on the corresponding supergroup manifold
as
\bea
\CL_{WZ}=\CC^{(6)}+\CF\CC^{(4)}+\half\CF^2\CC^{(2)},
\eea
where
\bea
 \CC^{(6)}=
\tC^{(6)}
+\frac{1}{3}\tB^{(2)}\tC^{(4)}
+\frac{1}{6}\tB^{(2)}\tB^{(2)}\tC^{(2)}
\eea
and
\bea
\tC^{(6)}&=&
-\frac{1}{90}
  \Pi^{a_1}\cdots\Pi^{a_5}\Pi_{a_1\cdots a_5}^{D}
+\frac{281}{13860}
  \Pi^{a_1}\cdots\Pi^{a_4}\Pi^\pmba\Pi_{a_1\cdots a_4\pmba}^{D}
+\frac{52}{3465}
  \Pi^{a_1}\Pi^{a_2}\Pi^{a_3}\Pi^\pmba\Pi^\pmbb\Pi_{a_1a_2a_3\pmba\pmbb}^{D}
\nonum\\&&
-\frac{47}{13860}
  \Pi^{a_1}\Pi^{a_2}\Pi^{\pmba_1}\Pi^{\pmba_2}\Pi^{\pmba_3}
  \Pi_{a_1a_2\pmba_1\pmba_2\pmba_3}^{D}
+\frac{1}{462}
  \Pi^{a}\Pi^{\pmba_1}\cdots\Pi^{\pmba_4}\Pi_{a\pmba_1\cdots\pmba_4}^{D}
+\frac{1}{6930}
  \Pi^{\pmba_1}\cdots\Pi^{\pmba_5}\Pi_{\pmba_1\cdots\pmba_5}^{D}
\nonum\\&&
+\Pi^a\Pi_a\Big[
    \frac{19}{1115}\Pi^a\Pi^b\Pi^\pmba\Pi_{ab\pmba}
    +\frac{2}{165}\Pi^a\Pi^\pmba\Pi^\pmbb\Pi_{a\pmba\pmbb}
    -\frac{19}{13860}\Pi^\pmba\Pi^\pmbb\Pi^\pmbga\Pi_{\pmba\pmbb\pmbga}
\Big]
\nonum\\&&
+\frac{1}{2}\Pi^\pmba\Pi_\pmba
\Big[
   \frac{52}{3465}\Pi^a\Pi^b\Pi^c\Pi_{abc}
   -\frac{3}{770}\Pi^a\Pi^b\Pi^\pmba\Pi_{ab\pmba}
   +\frac{4}{1155}\Pi^a\Pi^\pmba\Pi^\pmbb\Pi_{a\pmba\pmbb}
   -\frac{1}{1540}\Pi^\pmba\Pi^\pmbb\Pi^\pmbga\Pi_{\pmba\pmbb\pmbga}
\Big]
\nonum\\&&
+\frac{3}{385}\tB^{(2)}
(\tB^{(2)}\Pi^a{\Pi_a}'-\tC^{(2)}\Pi^a{\Pi_a})
-\frac{1}{231}\Pi^a{\Pi_a}
(\tB^{(2)}\Pi^b{\Pi_b}'-\tC^{(2)}\Pi^b{\Pi_b}).
\eea
In fact, one finds that
\bea
d\CC^{(6)}+\CC^{(4)}d\CF&=&
\frac{1}{30}\Pi^{a_1}\cdots\Pi^{a_5}\Pi^\pmba\Pi^\pmbb
(\ga_{a_1\cdots a_5}\sigma_1)_{\pmba\pmbb},
\eea
so that $d\CL_{WZ}=h_7$.
The supersymmetric WZ term is thus obtained,
and new coordinates are contained in a surface term,
due to the fiber bundle structure of our superalgebra.

For NS5-branes, we use the MC equations listed in the appendix B
in exchange for those for D5-branes.
The corresponding superalgebra is called the NS5-brane superalgebra.
We find that the supersymmetric WZ term for NS5-branes
can be constructed as
\bea
\hat\CL_{WZ}=
  -\CB^{(6)}
  +\CG\CB^{(4)}
  -\half\CG^2\CB^{(2)},
\eea
where
\bea
 \CB^{(6)}=
   \tB^{(6)}
   -\frac{1}{3}\tC^{(2)}\tC^{(4)}
   +\frac{1}{6}\tC^{(2)}\tC^{(2)}\tB^{(2)}
\eea
and
\bea
\tB^{(6)}&=&
-\frac{1}{90}
  \Pi^{a_1}\cdots\Pi^{a_5}\Pi_{a_1\cdots a_5}^{NS}
+\frac{281}{13860}
  \Pi^{a_1}\cdots\Pi^{a_4}\Pi^\pmba\Pi_{a_1\cdots a_4\pmba}^{NS}
+\frac{52}{3465}
  \Pi^{a_1}\Pi^{a_2}\Pi^{a_3}\Pi^\pmba\Pi^\pmbb\Pi_{a_1a_2a_3\pmba\pmbb}^{NS}
\nonum\\&&
-\frac{47}{13860}
  \Pi^{a_1}\Pi^{a_2}\Pi^{\pmba_1}\Pi^{\pmba_2}\Pi^{\pmba_3}
  \Pi_{a_1a_2\pmba_1\pmba_2\pmba_3}^{NS}
+\frac{1}{462}
  \Pi^{a}\Pi^{\pmba_1}\cdots\Pi^{\pmba_4}\Pi_{a\pmba_1\cdots\pmba_4}^{NS}
+\frac{1}{6930}
  \Pi^{\pmba_1}\cdots\Pi^{\pmba_5}\Pi_{\pmba_1\cdots\pmba_5}^{NS}
\nonum\\&&
-\Pi^a{\Pi_a}'\Big[
    \frac{19}{1155}\Pi^a\Pi^b\Pi^\pmba\Pi_{ab\pmba}
    +\frac{2}{165}\Pi^a\Pi^\pmba\Pi^\pmbb\Pi_{a\pmba\pmbb}
   -\frac{19}{13860}\Pi^\pmba\Pi^\pmbb\Pi^\pmbga\Pi_{\pmba\pmbb\pmbga}
\Big]
\nonum\\&&
-\frac{1}{2}\Pi^\pmba{\Pi_\pmba}'
\Big[
    \frac{52}{3465}\Pi^a\Pi^b\Pi^c\Pi_{abc}
   -\frac{3}{770}\Pi^a\Pi^b\Pi^\pmba\Pi_{ab\pmba}
   +\frac{4}{1155}\Pi^a\Pi^\pmba\Pi^\pmbb\Pi_{a\pmba\pmbb}
   -\frac{1}{1540}\Pi^\pmba\Pi^\pmbb\Pi^\pmbga\Pi_{\pmba\pmbb\pmbga}
\Big]
\nonum\\&&
+\frac{3}{385}\tC^{(2)}
(\tC^{(2)}\Pi^a{\Pi_a}-\tB^{(2)}\Pi^a{\Pi_a}')
-\frac{1}{231}\Pi^a{\Pi_a}'
(\tC^{(2)}\Pi^b{\Pi_b}-\tB^{(2)}\Pi^b{\Pi_b}').
\eea
In fact, one find that
\bea
-d\CB^{(6)}+\CB^{(4)}d\CF&=&
\frac{1}{30}\Pi^{a_1}\cdots\Pi^{a_5}\Pi^\pmba\Pi^\pmbb
(\ga_{a_1\cdots a_5}\sigma_3)_{\pmba\pmbb},
\eea
so that $d\hat\CL_{WZ}=\hat h_7$.
The supersymmetric WZ term is thus obtained,
and new coordinates are contained in a surface term.
In summary, we obtained the superinvariant WZ terms $\CL_{WZ}$
and $\hat \CL_{WZ}$ of D5- and NS5-branes respectively.
The total actions are identical to those found in sec. 2 up to
a surface term, and $\kappa$-invariant.

The relations of the superinvariant WZ terms of NS5- and D5-branes are as
follows.
$h^{(7)}$ and $\hat h^{(7)}$ are regarded as the pullback to the worldvolume
of the field strengths $\tH^{(7)}$ and $-\tG^{(7)}$ of the spacetime gauge
potentials $\tC^{(6)}$ and $\tB^{(6)}$
\bea
\tH^{(7)}=
d\tC^{(6)}-\frac{2}{3}\tC^{(4)}\tG^{(3)}+\frac{1}{3}\tB^{(2)}\tH^{(5)},
\qquad
\tG^{(7)}=
d\tB^{(6)}+\frac{2}{3}\tC^{(4)}\tH^{(3)}-\frac{1}{3}\tC^{(2)}\tH^{(5)}.
\label{H(7)G(7)}
\eea
The Bianchi identities are
\bea
d\tH^{(7)}-\tH^{(5)}\tG^{(3)}=0,\quad
d\tG^{(7)}+\tH^{(5)}\tH^{(3)}=0,
\eea
which correspond to $dh^{(7)}+h^{(5)}d\CF=0$ of (\ref{Dp;Bianchi})
and $d\hat h^{(7)}+\hat h^{(5)}d\CG=0$ of (\ref{NSp;Bianchi}).
The field strengths (\ref{H(7)G(7)})
are invariant under the spacetime gauge transformations:
(\ref{gauge tfn;1}), (\ref{gauge tfn;2}) and
\bea
\de\tC^{(6)}&=&
   d\Lambda^{(5)}
   +\frac{2}{3}d\Lambda^{(3)}\tB^{(2)}
   -\frac{1}{3}d\Omega^{(1)}\tC^{(4)}
   -\frac{1}{6}d\Omega^{(1)}\tB^{(2)}\tC^{(2)}
   +\frac{1}{6}d\Lambda^{(1)}\tB^{(2)}\tB^{(2)},
\label{gauge tfn;3;1}\\
\de\tB^{(6)}&=&
   d\Omega^{(5)}
   -\frac{2}{3}d\Lambda^{(3)}\tC^{(2)}
   +\frac{1}{3}d\Lambda^{(1)}\tC^{(4)}
   -\frac{1}{6}d\Lambda^{(1)}\tB^{(2)}\tC^{(2)}
   +\frac{1}{6}d\Omega^{(1)}\tC^{(2)}\tC^{(2)}.
\label{gauge tfn;3;2}
\eea
The gauge invariances of $I$ and $\hat I$
are assured by $\de h^{(7)}=0$ and $\de \hat h^{(7)}=0$,
which are guaranteed by the gauge transformations (\ref{gauge tfn;1}),
(\ref{gauge tfn;2}), (\ref{gauge tfn;3;1}) and (\ref{gauge tfn;3;2}).
Thus we have obtained the gauge invariant, $\kappa$-invariant and manifestly
superinvariant IIB D/NS5-brane action.

The WZ  terms $\CL_{WZ}$ for D5-branes and $\hat\CL_{WZ}$ for NS5-branes
are related by the S-duality:
(\ref{S-duality;1}), (\ref{S-duality;3}), (\ref{S-duality;wv}) and
\bea
\tB^{(6)}\rightarrow-\tC^{(6)},\quad
\tC^{(6)}\rightarrow \tB^{(6)}.
\eea
Under these transformations, the DBI actions $I_{DBI}$ and $\hat I_{DBI}$
are interchanged, and thus the obtained actions $I$ and $\hat I$ are
S-dual to each other.
The S-duality transformations are expressed as transformations on
the supercurrents as
\bea
\ba{r@{\,}c@{\,}lr@{\,}c@{\,}lr@{\,}c@{\,}l}
\Pi_{a_1\cdots a_5}^{NS}
  &\rightarrow&-\Pi_{a_1\cdots a_5}^{D},
&
\Pi_{a_1\cdots a_4\pmba}^{NS}
  &\rightarrow&-\Pi_{a_1\cdots a_4(\sigma\pmba)}^{D},
&
\Pi_{a_1\cdots a_3\pmba\pmbb}^{NS}
  &\rightarrow&-\Pi_{a_1\cdots a_3(\sigma\pmba)(\sigma\pmbb)}^{D},
\\
\Pi_{a_1a_2\pmba_1\cdots\pmba_3}^{NS}
  &\rightarrow&-\Pi_{a_1a_2(\sigma\pmba_1)\cdots(\sigma\pmba_3)}^{D},
&
\Pi_{a\pmba_1\cdots\pmba_4}^{NS}
  &\rightarrow&-\Pi_{a(\sigma\pmba_1)\cdots(\sigma\pmba_4)}^{D},
&
\Pi_{\pmba_1\cdots\pmba_5}^{NS}
  &\rightarrow&-\Pi_{a(\sigma\pmba_1)\cdots(\sigma\pmba_5)}^{D},
\\
\Pi_{a_1\cdots a_5}^{D}
  &\rightarrow&\Pi_{a_1\cdots a_5}^{NS},
&
\Pi_{a_1\cdots a_4\pmba}^{D}
  &\rightarrow&\Pi_{a_1\cdots a_4(\sigma\pmba)}^{NS},
&
\Pi_{a_1\cdots a_3\pmba\pmbb}^{D}
  &\rightarrow&\Pi_{a_1\cdots a_3(\sigma\pmba)(\sigma\pmbb)}^{NS},
\\
\Pi_{a_1a_2\pmba_1\cdots\pmba_3}^{D}
  &\rightarrow&\Pi_{a_1a_2(\sigma\pmba_1)\cdots(\sigma\pmba_3)}^{NS},
&
\Pi_{a\pmba_1\cdots\pmba_4}^{D}
  &\rightarrow&\Pi_{a(\sigma\pmba_1)\cdots(\sigma\pmba_4)}^{NS},
&
\Pi_{\pmba_1\cdots\pmba_5}^{D}
  &\rightarrow&\Pi_{a(\sigma\pmba_1)\cdots(\sigma\pmba_5)}^{NS}.
\ea
\label{S-duality;currents;5}
\eea
These transformations are shown to be the generalized R-symmetry,
and thus the S-duality is realized as an automorphism of our superalgebra,
as will be seen in the sec. 4.4 and 4.5.

\subsection{Reduction to N=(1,0)}
Our superalgebra contains N=(1,0) superalgebras as its subalgebra.

The F-string superalgebra corresponding to
(\ref{alg;supertranslation}) and (\ref{alg;F1})
is schematically described as
\bea
\{Q_\a,Q_\a\}\sim P^+,\quad
\{Q_\at,Q_\at\}\sim P^-,\quad
[Q_\a,P^+]\sim Z^\a,\quad
[Q_\at,P^-]\sim Z^\at,
\eea
where $P^\pm$ denotes $P_a\pm Z^a$.
The quotient algebra with respect to the ideal $\{Q_\at, P^-, Z^\at\}$
is the N=1 superalgebra found in \cite{Green}.
The WZ term for N=1 F-strings
is characterized by
$h^{(3)}=\Pi^a\Pi^\a\Pi^\b(\ga_a)_{\a\b}$.
One finds that the superinvariant WZ term is expressed as
$\CL_{WZ}=-\CB^{(2)}$
where $\CB^{(2)}=-\Pi^a\Pi_a-\half\Pi^\a\Pi_\a$.
This agrees with the one found in \cite{Siegel}.

Keeping only $Q_{\hat A}$, $Z^{\hat A}$ and $Z^{\hat A_1 \cdots \hat A_5}$,
where $\hat A$ collectively denotes $a$ and $\a$, in the NS5-brane
superalgebra, we obtain N=(1,0) NS5-brane superalgebra,
which is similar to those obtained in \cite{Sezgin2}
(with a trivial scaling of $\Pi_{\a_1\cdots\a_5}$).
The WZ term $\CL_{WZ}$ is characterized by
$h^{(7)}=\frac{1}{30}\Pi^{a_1}\cdots\Pi^{a_5}\Pi^\a\Pi^\b
(\ga_{a_1\cdots a_5})_{\a\b}$.
The superinvariant WZ term is $\CL_{WZ}=-\CB^{(6)}$
where
\bea
\CB^{(6)}&=&
   -\frac{1}{90}\Pi^{a_1}\cdots\Pi^{a_5}\Pi_{a_1\cdots a_5}^{NS}
   +\frac{281}{13860}\Pi^{a_1}\cdots\Pi^{a_4}\Pi^\a\Pi_{a_1\cdots a_4\a}^{NS}
   +\frac{52}{3465}\Pi^{a_1}\Pi^{a_2}\Pi^{a_3}\Pi^\a\Pi^\b\Pi_{a_1a_2a_3\a\b}
   ^{NS}
\nonum\\&&
   -\frac{47}{13860}\Pi^{a_1}\Pi^{a_2}\Pi^{\a_1}\Pi^{\a_2}\Pi^{\a_3}
       \Pi_{a_1a_2\a_1\a_2\a_3}^{NS}
   +\frac{1}{462}\Pi^a\Pi^{\a_1}\cdots\Pi^{\a_4}\Pi_{a\a_1\cdots\a_4}^{NS}
\nonum\\&&
   +\frac{1}{6930}\Pi^{\a_1}\cdots\Pi^{\a_5}\Pi_{\a_1\cdots\a_5}^{NS}.
\eea
The new coordinates in the WZ term are found to be contained in a surface term.

\section{The IIB-Brane Superalgebra}
In the previous section, we have introduced a set of superalgebras.
Let us derive and discuss these superalgebras 
with keeping the relations to the M-algebra in this section.

Foe this purpose,
we briefly describe the relation of branes in the type-IIB theory
to those in the M-theory.
Eleven-dimensional M-theory
contains 32 Majorana supercharges $Q_\a$.
The 528 components of the anti-commutator $\{Q_\a,Q_\b\}$ of the supercharges
are decomposed into generators for translations, M2-branes and M5-branes as
${\bf 528=11+55+462}$:
\bea
\ba{llc}
\mbox{ Translation}&P_\mu&{\bf 11}\\
\mbox{ M2-brane (M9-brane)}&Z^{\mu\nu}&{\bf 55}\\
\mbox{ M5-brane (MKK6-brane)}&Z^{{\mu_1\cdots\mu_5}}&{\bf 462}
\ea
\nonum
\eea
where $\mu,\nu,\dots$ are eleven-dimensional Lorentz indices.
In the parentheses,
we write the branes corresponding to charges with a time index.
One calls these branes M-branes collectively.

The IIA-branes in ten-dimensions are obtained by a dimensional reduction
of the M-branes with respect to eleven-th direction, say $\na$-direction.
The 32 components of Majorana supercharges
in eleven-dimensions are reduced to
16+16 Majorana-Weyl spinors $(Q_\a,Q_\ad)$
with the opposite chirality.
The IIA-branes are related to the M-branes as
\bea
\ba{lll}
\mbox{ Translation}&P_a&{\bf 9\ (1)}\\
\mbox{ D0-brane}      &P_\na&{\bf 1}\\
\mbox{ F-string (NS9-brane)}&Z^{\na a}&{\bf 9\ (1)}\\
\mbox{ D2-brane (D8-brane)}&Z^{ab}&{\bf 36\ (9)}\\
\mbox{ D4-brane (D6-brane)}&Z^{\na a_1\cdots a_4}&{\bf 126\ (84)}\\
\mbox{ NS5-brane (KK5-brane)}&Z^{a_1\cdots a_5}&{\bf 126\ (126)}
\ea
\nonum
\eea
where $a,b,\dots$ are ten-dimensional Lorentz indices.
The generators for translations and D0-branes originate from
those for translations in eleven-dimensions,
the generators for F-strings and D2-branes from those for M2-branes,
and the generators for D4-branes and NS5-branes from
those for M5-branes.

The IIB-branes are related to the above IIA-branes under a T-duality
transformation.
The type-IIB theory is chiral, which has 16+16 Majorana-Weyl supercharges
$(Q_\a, Q_\at)$ with the same chirality.
In order to change the chirality of a half of the IIA supercharges,
$Q_\ad$ are multiplied by a ten-dimensional $\ga$-matrix $\ga^\sh$,
where the $\sh$-direction is a space-like direction
with respect to which a T-duality is performed.
One finds that
the obtained algebra is rewritten in a ten-dimensional covariant form
under the following identifications of the IIB-brane charges
and the M-brane charges:
\bea
\ba{llclc}
\mbox{ Translation}&P_a&=&P_i\oplus Z^{\na\sh}&{\bf 10}\\
\mbox{ F-string (NS9-brane)}&Z^a&=&Z^{\na i}\oplus P_\sh&{\bf 9\ (1)}\\
\mbox{ D-string (D9-brane)}&\S^a&=&Z^{\sh i}\oplus P_\na&{\bf 9\ (1)}\\
\mbox{ D3-brane (NS/D7-brane)}&\S^{abc}&=&Z^{\na\sh ijk}\oplus Z^{ij}&{\bf 84\
(36)}\\
\mbox{ D5-brane}&\S^{m_1\cdots m_5}&\sim&Z^{\na 0p_1p_2p_3}\oplus
         Z^{\na p_1\cdots p_4}&{\bf 126}\\
\mbox{ NS5-brane}&Z^{m_1\cdots m_5}&\sim&Z^{\sh p_1\cdots p_4}\oplus
         Z^{0\sh p_1p_2p_3}&{\bf 126}\\
\mbox{ KK5-brane}&P^{0m_1\cdots m_4}&\sim&Z^{0p_1\cdots p_4}\oplus
         Z^{p_1\cdots p_5}&{\bf 126}
\ea
\nonum
\eea
where $m$ runs all of the nine spacelike directions,
but $p$ runs except for the $\sh$-direction.
$i$ runs ten spacetime directions
except for the $\sh$-direction.
The D5-, NS5- and KK5-brane charges are self-dual.
The ``$\sim$'' means the equivalence up to the self-duality relation.
It follows that 
the S-duality in the IIB-branes is rephrased as a modular transformation
on $T^{2}$ spanned by $(\na,\sh)$.

\subsection{M-algebra, IIA-brane superalgebra
and T-duality}\label{M-algebra}
The ordinary superspace,
which is the supertranslation group
(``superPoincar\'e''/``Lorentz'')
generated by $Q_A=( P_a,Q_\pmba)$
is regarded as a generalization of the spacetime
which is the translation group
(``Poincar\'e''/``Lorentz'')
generated by $P_a$
(Lorentz group is an automorphism of the algebra).
We now extend the ordinary superspace
to those spanned by generators of $p$-branes $Z^{A_1\cdots A_p}$
as well as $Q_A$.
In order to consider the relations of the IIB-brane superalgebra
to the M-algebra,
we begin with describing the M-algebra \cite{Sezgin},
which is an eleven dimensional superalgebra
generated by\footnote{
We omit ``superstring'' contained in the M-algebra.
}
\bea
\ba{llcl}
\mbox{ Supertranslation}&Q_M&\Leftrightarrow&
   \Pi^M\\
\mbox{ M2-brane (M9-brane)}&Z^{MN}&\Leftrightarrow&
   \Pi_{MN}\\
\mbox{ M5-brane (MKK6-brane)}&Z^{M_1\cdots M_5}&\Leftrightarrow&
   \Pi_{M_1\cdots M_5}
\ea
\nonum
\eea
where eleven-dimensional Lorentz indices $\mu$
and 32-component Majorana spinor indices $\a$
are collectively denoted as  $M,N,\cdots$,
so that $Q_M=(P_\mu,Q_\a)$, $Z^{MN}=(Z^{\mu\nu},Z^{\mu\a},Z^{\a\b})$, etc.
The right hand sides of arrows denote the corresponding LI one-form
on the supergroup manifold.

The IIA-brane superalgebra is obtained by
a dimensional reduction of this M-algebra to ten-dimensions,
and generated by
\bea
\ba{llcrcl}
\mbox{ Supertranslation}
   &Q_A&\Leftrightarrow
   &\Pi^A&=&(\Pi^a,\Pi^\a,\Pi^\ad)
\\
\mbox{ F-string (NS9-brane)}
   &Z^A&\Leftrightarrow
   &\Pi_A&=&(\Pi_{\na a},\Pi_{\na\a},\Pi_{\na\ad})\\
\mbox{ D0-brane}
   &\S&\Leftrightarrow
   &\Pi&=&(\Pi^\na)\\
\mbox{ D2-brane (D8-brane)}
   &\S^{AB}&\Leftrightarrow
   &\Pi_{AB}&=&(\Pi_{ab},\cdots)\\
\mbox{ D4-brane (D6-brane)}
   &\S^{A_1\cdots A_4}&\Leftrightarrow
   &\Pi_{A_1\cdots A_4}&=&(\Pi_{\na a_1\cdots a_4},\cdots)\\
\mbox{ NS5-brane (KK5-brane)}
   &Z^{A_1\cdots A_5}&\Leftrightarrow
   &\Pi_{A_1\cdots A_5}&=&(\Pi_{a_1\cdots a_5},\cdots)
\ea
\nonum
\eea
where
the indices $A,B,\cdots$ collectively denote ten-dimensional Lorentz indices
$a$
and
16-component Majorana-Weyl spinor indices $\a$ and $\ad$
with positive and negative chirality, respectively.
Branes corresponding to the generators with a time index
are listed in parentheses.
Dots denote the forms with Lorentz indices exchanged for spinorial indices
of the corresponding forms.

In order to obtain the type-IIB theory, we must take
a  T-dual
with respect to a space-like direction $x^\sh$.
As a T-duality transformation (see \cite{Townsend lecture 1997}),
we perform a dimensional reduction to nine-dimensions
with respect to $x^\sh$
and
change the chirality of dotted spinors by multiplying $\ga^\sh$ as:
$\Pi_{a\at}={(\ga^\sh)^\bd}_\at\Pi_{a\bd}$ etc.

One may apparently expect that the IIB-brane superalgebra
would be completely obtained
by a T-duality
from the IIA-brane superalgebra.
We find however that the generators of supertranslations, F-strings,
D-strings
and D3-branes originate from those of the IIA-brane superalgebra,
but some charges of IIB D5- and IIB NS5-branes
are not obtained by a T-duality as will be seen below.
\subsection{Supertranslations, F-strings and  D-strings}
The superalgebra in presence of F-strings and/or D-strings
are related to those in presence of F-strings and/or
D0- and D2-branes in the
IIA superalgebra
under a T-duality.
This is found to be rewritten in a ten-dimensional covariant form
under the identification \cite{sakagu1}:
\bea
\ba{lrcl}
\mbox{ Supertranslation}&Q_A&\Leftrightarrow&
  \Pih^a=(\Pi^i,\Pi_{\na\sh}),\
  \Pih^\pmba=\Pi^\pmba,
\\
\mbox{ F-string}&Z^A&\Leftrightarrow&
  \Pih_a=(\Pi_{\na i}, \Pi^\sh),\
  \Pih_\pmba=\Pi_{\na\pmba},
\\
\mbox{ D-string}&\S^A&\Leftrightarrow&
  {\Pih_a}'=(\Pi_{\sh i},-\Pi^\na),\
  {\Pih_\pmba}'=\Pi_{\sh\pmba}.
\ea
\label{charges;1}
\eea
The F-string superalgebra is generated by $(Q_A, Z^A)$
and the corresponding MC equations are
(\ref{alg;supertranslation}) and (\ref{alg;F1}).
The D-string superalgebra is generated by $(Q_A, \S^A)$
and the corresponding MC equations are
(\ref{alg;supertranslation}) and (\ref{alg;D1}).
We denote these superalgebras as $sP[Z^A]$ and $sP[\S^A]$, respectively.
The Jacobi identities, or equivalently the integrability conditions $d^2=0$,
are guaranteed by the Fierz identities:
$(\ga_a)_{(\a\b}(\ga^a)_{\ga\de)}=0$ for F-strings,
and
$(\ga_a)_{\at(\b}(\ga^a)_{\ga\de)}=0$ for D-strings,
which are encountered in (\ref{FDA2;Fierz;F1}) and (\ref{FDA1;Fierz;D1})
in the CE cohomology classification
of IIB-branes.

The S-duality
is expressed as a modular transformation on $T^2$ spanned by
$(\na,\sh)$.
Now, this generalizes to include new fermionic generators
$\hat\Pi_\pmba$ and $\hat\Pi_\pmba'$.
One finds that
under the S-duality transformation (\ref{S-duality;currents;1}),
the spacetime superalgebra (\ref{alg;F1}) interchanges with (\ref{alg;D1})
and (\ref{alg;supertranslation}) stays invariant.

The superalgebra generated by generators listed
in (\ref{charges;1}), denoting $sP[Z^A,\S^A]$,
contains various subalgebras.
Noting that new fermionic generators are the centers
of the algebra $sP[Z^A,\S^A]$,
and that $Z^a$ and $\S^a$ are the centers of the algebra $sP[Z^a,\S^a]$,
one finds
\bea
sP[Z^A,\S^A]\supset
\left\{ \ba{c}
sP[{Z^\a Z^\at\choose Z^a},{\S^\a \hs{3}\choose \S^a}]\\
\vdots\\
sP[{\hs{3} Z^\at\choose Z^a},{\S^\a \S^\at \choose \S^a}]
\ea\right.
\supset\cdots\supset
sP[Z^a,\S^a] \supset
{sP[Z^a] \atopwithdelims\{. sP[\S^a] } \supset sP,
\eea
where $sP$ is generated by $Q_A$ as (\ref{alg;supertranslation}).
The superalgebra $sP[Z^A,\S^A]$ contains $sP[Z^A]$ or $sP[\S^A]$
as a subalgebra since $Z^A$ and $\S^A$ are ideals of the algebra
$sP[Z^A,\S^A]$
\bea
sP[Z^A,\S^A]\supset
\left\{
\ba{l}
sP[Z^A]\supset
\left\{ \ba{l} sP[{Z^\a\hs{3}\choose Z^a}]\\ sP[{\hs{3}Z^\at\choose Z^a}]
\ea\right.
  \supset sP[Z^a],\\
sP[\S^A] \supset
\left\{ \ba{l} sP[{\S^\a\hs{3}\choose\S^a}] \\ sP[{\hs{3}\S^\at\choose\S^a}]
\ea\right.
  \supset sP[\S^a],
\ea
\right.
\supset sP.
\eea
Due to the property that the superalgebras contain $sP$ as a subalgebra,
coordinates corresponding to the generators $Z^A$ and $\S^A$
are not contained in supercurrents $\Pi^A$.
The fiber bundle structure with the base space $sP$
is a universal structure of our superalgebras.

\subsection{D3-branes}
The D3-brane charges with three bosonic indices are related to
those of the wrapped D4-brane and unwrapped D2-brane in the IIA superalgebra,
which are $Z^{p_1p_2}$ and $Z^{\na\sh p_1p_2p_3}$ in the M-algebra,
where $p_i$ runs 1 to 9 except for $\sh$.
We find that the superalgebra obtained by a T-duality
from the IIA D2- and D4-brane superalgebra
is rewritten in a ten-dimensional covariant form
and the D3-brane charges $\S^{ABC}$ are completely determined
in terms of the charges in the M-algebra:
\bea
\ba{lcl}
\mbox{ D3-brane}\ \S^{ABC} &\Leftrightarrow&
\left\{
\ba{l}
\Pih_{abc}=
	(\Pih_{ijk}=\Pi_{\na\sh ijk}, \Pih_{\sh ij}=-\Pi_{ij}),\\
\Pih_{abb\pmba}=
	(\Pih_{ij\pmba}=\Pi_{\na\sh ij\pmba}, \Pih_{\sh i\pmba}=\Pi_{i\pmba}),
\\
\Pih_{a\pmba\pmbb}=
	(\Pih_{i\pmba\pmbb}=\Pi_{\na\sh i\pmba\pmbb},
        \Pih_{\sh\pmba\pmbb}=-\Pi_{\pmba\pmbb}),
\\
\Pih_{\pmba\pmbb\pmbga}=\Pi_{\na\sh\pmba\pmbb\pmbga}.
\ea\right.
\ea
\label{charges;3}
\eea
The S-duality, which is a modular transformation on $T^2$ spanned by
$(\na,\sh)$,
generalizes to include new fermionic generators
$\S^{ab\pmba}$, $\S^{a\pmba\pmbb}$ and $\S^{\pmba\pmbb\pmbga}$.
We obtain the D3-brane superalgebra (dropping hats)
found in (\ref{alg;D3}).
Jacobi identities are satisfied due to
\bea
({\ga^{ab}}_c)_{\at(\b}(\ga^c)_{\ga\de)}
+2(\ga^{[a})_{\at(\b}(\ga^{b]})_{\ga\de)}
=0,
\eea
which is the characteristic identity of D3-branes,
and encountered in (\ref{FDA1;Fierz;D3}) in the CE cohomology classification
of IIB-branes.

The S-duality was expressed as a set of transformations of supercurrents:
(\ref{S-duality;currents;1}) and (\ref{S-duality;currents;3}),
which is a generalization of the R-symmetry.
We find that these transformations leave
the MC equations (\ref{alg;D3}) for D3-branes invariant.
The S-duality associated with the generalized R-symmetry is realized as
an automorphism of our algebra.

Interestingly, we find that the existence of the D3-brane charges $Z^{ABC}$
in the superalgebra necessarily requires the presence of the
F-string charges $Z^A$ and the D-string charges $\S^A$,
while the N=1 $p$-brane superalgebra \cite{BS}
is generated by supertranslation and $p$-brane charges.
This is a characteristic feature of our superalgebras.
This is consistent with the fact that
the WZ terms for D3-branes are composed of the
spacetime gauge potentials which couple to D3-branes, D-strings and F-strings,
or that the gauge field strength of the four-form gauge potential
which couples to D3-branes
contains the gauge potentials which couple to D-strings and F-strings.

The D3-brane superalgebra contains the F- and D-string superalgebras
as subalgebras.
In addition, the D3-brane superalgebra contains various subalgebras.
We assign a degree $(p;q,\tilde q)$ to the dual form
$\Pi_{a_1\cdots a_p\a_1\cdots\a_q\bt_{1}\cdots\bt_{\tilde q}}$
of generators.
Under an operation of the differential $d$, the
forms with degree $(p;q,\tilde q)$
transform to those with degree:
\bea
&
(p+2;q-2,\tilde q),\quad (p+2;q,\tilde q-2),
&\nonum\\&
(p+1;q-2,\tilde q+1),\quad(p+1;q-1,\tilde q),\quad
(p+1;q,\tilde q-1),\quad (p+1;q+1,\tilde q-2),
&\label{deg;d}
\eea
and those with degree $(p';q',\tilde q')$ where $p'+q'+\tilde q'=1$.
In order to close an algebra,
Jacobi identities, or equivalently $d^2=0$, must be satisfied.
A subalgebra is generated by
the forms with degree $(p;q,\tilde q)$,
those with degree (\ref{deg;d})
and those obtained by differentiating the forms with degree (\ref{deg;d}).
The new generators $\S^{\pmba\pmbb\pmbga}$ are the centers of
the D3-brane superalgebra $sP[Z^A,\S^A,\S^{ABC}]$,
and the new generators $\S^{a\pmba\pmbb}$ are the centers of
the subalgebra $sP[Z^A,\S^A,\S^{aAB}]$, and so on.
In addition, $Z^\pmba$, $\S^\pmba$ and $\S^{abc}$ are center of the algebra
$sP[Z^A,\S^A,\S^{abc}]$.
In general, letting $\mathfrak g$ be a Lie algebra,
the sequences $\mathfrak h^0, \mathfrak h^1, \cdots$ defined by
$
\mathfrak h^0\equiv \mathfrak g,
\mathfrak h^1\equiv [\mathfrak g,\mathfrak h],
\mathfrak h^2\equiv [\mathfrak g,\mathfrak h^1],\cdots,
\mathfrak h^n\equiv [\mathfrak g,\mathfrak h^{n-1}],\cdots,
$
is called the descending central sequence.
These satisfy
$
\mathfrak h^0\supset
\mathfrak h^1\supset
\mathfrak h^2\supset\cdots\supset
\mathfrak h^n\supset\cdots
$
and $\mathfrak h^0, \mathfrak h^1, \mathfrak h^2,\cdots$ are ideals
in $\mathfrak g$.
In our case, the descending central sequence is
$\mathfrak h^0=sP[Z^A,\S^A,\S^{ABC}]$,
$\mathfrak h^1=\{P_a, Z^A, \S^A,\S^{ABC} \}$,
$\mathfrak h^2=\{Z^\pmba, \S^\pmba, \S^{\pmba AB} \}$,
$\mathfrak h^3=\{\S^{\pmba\pmbb A} \}$,
$\mathfrak h^4=\{\S^{\pmba\pmbb\pmbga} \}$ and
$\mathfrak h^5=\{0 \}$.
Our algebra is nilpotent and
ideals $\mathfrak h^1$, $\mathfrak h^2$, $\mathfrak h^3$ and $\mathfrak h^4$
are said to be proper.
Especially, $\mathfrak h^4$ is abelian ideal, said to be
the center in $sP[Z^A,\S^A,\S^{ABC}]$.
For a given ideal $I$, the quotient Lie algebra is defined
by ${\mathfrak g}/\sim$, where $\sim$ denotes the equivalent relation:
$x\sim x+I$ for $x\in \mathfrak g$.
The quotient Lie algebra $\mathfrak g/\sim$ is a subalgebra of $\mathfrak g$.
It follows that the D3-brane superalgebra contains various
subalgebras:
\bea
&&
sP[Z^A,\S^A,\S^{ABC}]\supset
\left\{\ba{c}
sP[Z^A,\S^A,{\S^{\a\b\ga}\S^{\a\b\gat}\S^{\a\bt\gat}\hs{5}
\choose \S^{ABa} }]\\
\vdots\\
sP[Z^A,\S^A,{\hs{5}\S^{\a\b\gat}\S^{\a\bt\gat}\S^{\at\bt\gat}
\choose \S^{ABa} }]
\ea\right.
\supset\cdots\supset
sP[Z^A,\S^A,\S^{ABa}]\supset
\nonum\\&&
\supset
\left\{\ba{c}
sP[Z^A,\S^A,{\S^{a\a\b}\S^{a\a\bt}\hs{6}\choose \S^{Aab}}]\\
\vdots\\
sP[Z^A,\S^A,{\hs{6}\S^{a\a\bt}\S^{a\at\bt}\choose \S^{Aab}}]
\ea\right.
\supset\cdots\supset
sP[Z^A,\S^A,\S^{Aab}]\supset
\left\{ \ba{c}
sP[Z^A,\S^A,{\S^{ab\a}\hs{6} \choose \S^{abc}}]\\
sP[Z^A,\S^A,{\hs{6}\S^{ab\at} \choose \S^{abc}}]
\ea\right.
\supset\nonum\\&&
\supset sP[Z^A,\S^A,\S^{abc}]
\supset
\left\{\ba{l}
sP[Z^A,\S^A]\\
\supset\cdots\supset sP[Z^a,\S^a,\S^{abc}]
\ea\right.
\supset\cdots\supset sP.
\eea

\subsection{D5-branes}
The D5-brane is related to the unwrapped D4-brane
and the wrapped D6-brane
under a T-duality.
The 126 D5-brane charges with five bosonic charges
$\S^{m_1\cdots m_5}, m_i=1,\cdots ,9$,
are obtained from the 70 D4-brane charges $Z^{\na p_1\cdots p_4}$
and the 56 D6-brane charges $Z^{\na 0 p_1p_2 p_3}$,
as $\S^{\sh p_1\cdots p_4}=Z^{\na p_1\cdots p_4}$
and $\S^{p_1\cdots p_5}\sim Z^{\na 0p_1p_2p_3}$,
where ``$\sim$'' denotes the equivalence modulo the self-duality relation.
Thus, the charges naturally related to those in the IIA superalgebra are not
$\S^{m_1\cdots m_5}$ but $\S^{\na i_1\cdots i_4}$.
Under the identification of the D5-brane charges with the charges in the
M-algebra,
\bea
\ba{lcl}
\mbox{ D5-brane}\ \S^{\sh I_1\cdots I_4}&\Leftrightarrow&
\Pih_{\sh I_1\cdots I_4}^{D}=-\Pi_{\na I_1\cdots I_4},
\ea
\label{charges;D5}
\eea
where $I$ collectively denotes $i$ and $\pmba$,
we find that the resultant superalgebra is rewritten
in a ten-dimensional covariant form.
For example, the MC equation
\bea
d\Pih_{\sh i_1i_2i_3\a}^{D}&=&
   -\Pi^\bt\Pih^a(\ga_{a\sh i_1i_2i_3})_{\a\bt}
   -\Pi^\b\Pih_{a\sh i_1i_2i_3}^{D}(\ga^a)_{\a\b}
\nonum\\&&
   -4\Pi^\bt\Pih_{[\sh}(\ga_{i_1i_2i_3]})_{\a\bt}
   +4\Pi^\b\Pih_{[\sh i_1i_2}(\ga_{i_3]})_{\a\b}
\eea
is obtained from the IIA D4-brane superalgebra,
where hatted forms are defined in (\ref{charges;1}), (\ref{charges;3})
and (\ref{charges;D5}).
The corresponding Jacobi identity is found to be satisfied by a
$(abcd)=(\sh i_1\cdots i_4)$ part of the Fierz identity
\bea
(\phi\ga_e\phi)(\psi\ga^{abcde}\phi)-4(\phi\ga^{[a}\phi)(\psi\ga^{bcd]}\phi)=0,
\label{Fierz;D5}
\eea
which is appeared in (\ref{FDA1;Fierz;D5})
in the CE cohomology classification of IIB-branes.
It follows from this fact that the obtained superalgebra can be lifted into
a ten-dimensional covariant form.
In addition, the superalgebra can be expressed in terms of
SO(2) doublets of the supercurrents:
$\Pi_{a_1\cdots a_4\pmba}^{D}
={\Pi_{a_1\cdots a_4\at}^{D} \choose \Pi_{a_1\cdots a_4\a}^{D}}$
etc.
As a result, we obtain the MC equations for the D5-brane charges
(\ref{alg;D5-1})$\sim$(\ref{alg;D5-5})
in the appendix A.

In order to construct the supersymmetric form of the WZ term for D5-branes,
we need the charges with five spinorial indices $\S^{\pmba_1\cdots\pmba_5}$.
However, the charges with five spinorial indices obtained by a T-duality
from the IIA superalgebra, or equivalently the M-algebra,
turn out to be the IIB KK5-brane charges, as will be seen in sec.\ref{KK5}.
We here introduce the corresponding generators to the algebra for D5-branes.
Let us note that under an operation of a differential $d$,
the degree $(p;q, \tilde q)$ of the form of the D5-brane superalgebra
is transformed into a set of the D5-brane charges with degree (\ref{deg;d}).
This is a universal property of our superalgebra:
F-string, D-string, D3-brane and D5-brane.
The D5-brane superalgebra supplied with
$\S^{\pmba_1\cdots\pmba_5}$
turns out to possess this property too.
This property helps us to construct the MC equations for
the charges with five spinorial indices.
We can construct the Maurer-Cartan equation (\ref{alg;D5-6}).
We find again that the existence of the D5-brane charges
necessarily requires the existence of the charges of
superalgebra generated by supertranslations
F-strings, D-strings and D3-branes.
This is consistent with the form
of the WZ term for D5-branes,
or of the seven-form field strength $\tH^{(7)}$
of the six-form R$\otimes$R gauge potential $\tC^{(6)}$.

The D5-brane superalgebra contains various subalgebras.
The descending central series are
$\mathfrak h^0=
   sP[ Z^A, \S^A, \S^{ABC}, \S^{A_1\cdots A_5} ]$,
$\mathfrak h^1=
   \{ P_a, Z^A, \S^A,\S^{ABC}, \S^{A_1\cdots A_5} \}$,
$\mathfrak h^2=
   \{ Z^\pmba, \S^\pmba, \S^{\pmba AB}, \S^{\pmba A_1\cdots A_4} \}$,
$\mathfrak h^3=
   \{ \S^{\pmba\pmbb A}, \S^{\pmba\pmbb A_1A_2A_3} \}$,
$\mathfrak h^4=
   \{ \S^{\pmba\pmbb\pmbga}, \S^{\pmba\pmbb\pmbga A_1A_2} \}$,
$\mathfrak h^5=
   \{ \S^{\pmba_1\cdots\pmba_4 A} \}$,
$\mathfrak h^6=
   \{ \S^{\pmba_1\cdots\pmba_5} \}$,
$\mathfrak h^7=
   \{ 0 \}$.
These are the ideals of our algebra, and
$\mathfrak h^6$ is the center of $sP[Z^A,\S^A,\S^{ABC},\S^{A_1\cdots A_5}]$.
The generators $\S^{a\pmba_1\cdots\pmba_4}$
are the centers of the subalgebra
$sP[Z^A,\S^A,\S^{ABC},\S^{aA_1\cdots A_4}]$,
and so on.
It follows from this fact that
\bea
&&
sP[Z^A,\S^A,\S^{ABC},\S^{A_1\cdots A_5}]\supset
\left\{\ba{c}
sP[Z^A,\S^A,\S^{ABC},{\S^{\a_1\cdots \a_5}\cdots \S^{\a\bt_1\cdots \bt_4}\hs{7}
\choose \S^{A_1\cdots A_4a}}]\\
\vdots\\
sP[Z^A,\S^A,\S^{ABC}{\hs{7}\S^{\a_1\cdots \a_4\bt}\cdots \S^{\bt_1\cdots \bt_5}\choose \S^{A_1\cdots A_4a}}]\\
\ea\right.
\supset\cdots\supset\nonum\\&&
\supset
sP[Z^A,\S^A,\S^{ABC},\S^{A_1\cdots A_4a}]
\supset\cdots\supset
sP[Z^A,\S^A,\S^{ABC},\S^{a_1\cdots a_5}]
\supset\nonum\\&&
\supset\left\{\ba{l}
sP[Z^A,\S^A,\S^{ABC}]\\
\supset\cdots\supset sP[Z^a,\S^a,\S^{abc},\S^{a_1\cdots a_5}]
\ea\right.
\supset\cdots\supset sP.
\label{subalgebras;D5}
\eea

\subsection{NS5-branes}
As is well known,
under a T-duality:
$R\leftrightarrow l^2_s/R$ and $g_s\leftrightarrow l_sg_s/R$,
where $g_s$ and $l_s$ denote
the string coupling and the string length, respectively,
and $R$ is the radius of the $\sh$-direction,
the tension $R/g_s^2l_s^6$ of the wrapped IIA NS5-brane is transformed into
the tension $R/g_s^2l_s^6$ of the wrapped IIB NS5-brane.
The tension $R^2/g_s^2l_s^8$ of the IIA KK5-brane,
with the $R$ identified with
the radius of the $S^1$ of the Taub-NUT geometry,
is transformed into the tension $1/g_s^2l_s^6$
of the unwrapped IIB NS5-brane.
The 126 IIB NS5-brane charges with five bosonic indices $Z^{m_1\cdots m_5}$
are obtained from to the 70 IIA NS5-brane charges
$Z^{\sh p_1\cdots p_4}$ and
the 56 IIA KK5-brane charges $Z^{\sh 0p_1p_2p_3}$,
as $Z^{\sh p_1\cdots p_4}=Z^{\sh p_1\cdots p_4}$
and $Z^{p_1\cdots p_5}\sim Z^{\sh 0p_1p_2p_3}$,
where ``$\sim$'' denotes the equivalence modulo the self-duality relation.
The charges naturally related to those in the IIA superalgebra are not
$Z^{m_1\cdots m_5}$ but $Z^{\sh i_1\cdots i_4}$.
Under the identification of the charges of the M-algebra
with those in NS5-brane superalgebra:
\bea
\mbox{ NS5-brane}\ Z^{\sh I_1\cdots I_4}&\Leftrightarrow&
\hat \Pi_{\sh I_1\cdots I_4}^{NS}=\Pi_{\sh I_1\cdots I_4}.
\label{charges;NS5}
\eea
we find that the algebra is lifted into a ten-dimensional
covariant form.
For example, the equation
\bea
d\hat\Pi_{\sh i_1i_2i_3\a}^{NS}&=&
   \Pi^\b\Pih^a(\ga_{a\sh i_1i_2i_3})_{\a\b}
   +\Pi^\b\Pih_{a\sh i_1i_2i_3}^{NS}(\ga^a)_{\a\b}
\nonum\\&&
   -4\Pi^\bt{\Pih_{[\sh}}'(\ga_{i_1i_2i_3]})_{\a\bt}
   +4\Pi^\bt\Pih_{[i_1i_2}(\ga_{i_3]})_{\a\bt}
\eea
is obtained by a T-duality, where hatted forms are defined
in (\ref{charges;1}), (\ref{charges;3}) and (\ref{charges;NS5}).
The corresponding Jacobi identities are satisfied by $(abcd)=(\sh i_1i_2i_3)$
part of the Fierz identity in ten-dimensions
\bea
(\phi\ga_e\phi)(\psi\ga^{abcde}\psi)
-(\psi\ga_e\psi)(\phi\ga^{abcde}\phi)
+16(\phi\ga^{[a})(\phi\ga^{bcd]}\psi)=0,
\eea
which is (\ref{FDA2;Fierz;NS5})
in the CE cohomology classification of IIB-branes.
It follows from this fact that the superalgebra can be lifted
into a ten-dimensional covariant form.
The resulting MC equations are found to be expressed
in terms of SO(2) doublets
of the supercurrents:
$
\Pi^{NS}_{a_1\cdots a_4\pmba}=
{\Pi^{NS}_{a1\cdots a_4\at}\choose \Pi^{NS}_{a1\cdots a_4\a}}
$
as the NS5-brane superalgebra listed in (\ref{alg;NS5-1})
$\sim$(\ref{alg;NS5-5})
in the appendix B.

In order to construct the supersymmetric WZ term for NS5-branes,
we supply the superalgebra with the charges
with five spinorial indices $Z^{\pmba_1\cdots\pmba_5}$.
Noting that the MC equations for NS5-branes
(\ref{alg;NS5-1})$\sim$(\ref{alg;NS5-5})
possess the property that the charges with degree $(p;q,\tilde q)$
transform into those with degree (\ref{deg;d}),
we find the MC equation (\ref{alg;NS5-6})
in the appendix B.

We find that the existence of the NS5-brane charges in the superalgebra
necessarily requires the presence of the charges of
F-strings, D-strings and D3-branes.
This is consistent with the form of the WZ term for NS5-branes,
or of the seven-form field strength $\tG^{(7)}$ of the six-form NS$\otimes$NS
gauge potential $\tB^{(6)}$.
The NS5-brane superalgebra contains various subalgebras,
which is expressed as (\ref{subalgebras;D5}) with replacing
the D5-brane charges $\S^{A_1\cdots A_5}$
with the NS5-brane charges $Z^{A_1\cdots A_5}$.

The S-duality was given by a set of transformations
of supercurrents (\ref{S-duality;currents;1}),
(\ref{S-duality;currents;3})
and (\ref{S-duality;currents;5}).
We can show that under these transformations
the NS5-brane superalgebra is
mapped to the D5-brane superalgebra,
and thus the S-duality is realized as an automorphism of
$sP[Z^A, \S^A, \S^{ABC}, Z^{A_1\cdots A_5}, \S^{A_1\cdots A_5}]$.

\subsection{KK5-branes}\label{KK5}
Under a T-duality,
the tension $1/g_s^2l_s^6$ of the unwrapped IIA NS5-brane is transformed into
the tension $R^2/g_s^2l_s^8$ of the IIB KK5-brane with the $R$
identified with the radius $R_{TN}$ of the $S^1$ in the Taub-NUT geometry.
The tension $RR_{TN}^2/g_s^2l_s^8$ of the wrapped IIA KK5-brane
is transformed into the tension $RR^2_{TN}/g_s^2l_s^8$
of the wrapped IIB KK5-brane.
The 126 IIB KK5-brane charges with five bosonic indices $Z^{0m_1\cdots m_4}$
are related to the 56 IIA NS5-brane charges $Z^{p_1\cdots p_5}$ and
the 70 IIA KK5-brane charges $Z^{0p_1\cdots p_4}$
as $Z^{0\sh p_1p_2p_3}\sim Z^{p_1\cdots p_5}$
and $Z^{0p_1\cdots p_4}=Z^{0 p_1\cdots p_4}$,
where ``$\sim$'' denotes the equivalence modulo the self-duality relation.
The charges naturally related to those in the IIA superalgebra are not
$Z^{0m_1\cdots m_4}$ but $Z^{i_1\cdots i_5}$.
As was mentioned before, the algebra obtained by a T-duality
is not rewritten in a covariant form.
The MC equations for the KK5-brane charges are found in the appendix C.
The identification of the IIB KK5-brane charges with the M5-brane charges
are
\bea
\mbox{ KK5-brane}\ P^{I_1\cdots I_5}&\Leftrightarrow&
\Pi^{KK}_{I_1\cdots I_5}=\Pi_{I_1\cdots I_5}.
\eea
The characteristic identity\footnote{
Summing up with respect to $\sh=0,\cdots,9$ except for $i_1,\cdots,i_4$,
the covariant identity is obtained as
\bea
(\phi\ga_e\phi)(\psi\ga^{abcde}\psi)
+(\psi\ga_e\psi)(\phi\ga^{abcde}\phi)
+(\phi\ga_e\psi)(\psi\ga^{abcde}\phi)
+3(\phi\ga^{e[ab}\psi)(\psi{\ga_e}^{cd]}\phi)=0.
\nonum
\eea}
for the KK5-branes is found to be
\bea
&
(\ga^j1)_{\pmba\pmbb}(\ga_{ji_1\cdots i_4}1)_{\pmbga\pmbde}
+(\ga^\sh\sigma_3)_{\pmba\pmbb}(\ga_{\sh i_1\cdots i_4}\sigma_3)_{\pmbga\pmbde}
+(\ga^\sh\sigma_1)_{\pmba\pmbb}(\ga_{\sh i_1\cdots i_4}\sigma_1)_{\pmbga\pmbde}
&\nonum\\&
-3(\ga_{\sh i_1i_2}i\sigma_2)_{\pmba\pmbb}
   (\ga_{\sh i_3i_4}i\sigma_2)_{\pmbga\pmbde}=0.
&
\eea
Under the S-duality transformation of supercurrents:
(\ref{S-duality;currents;1}), (\ref{S-duality;currents;3}),
(\ref{S-duality;currents;5}) and
\bea
&
\Pi^{KK}_{i_1\cdots i_5}\rightarrow
   \Pi^{KK}_{i_1\cdots i_5},\quad
\Pi^{KK}_{i_1\cdots i_4\pmba}\rightarrow
   \Pi^{KK}_{i_1\cdots i_4(\sigma\pmba)},\quad
\Pi^{KK}_{i_1i_2i_3\pmba_1\pmba_2}\rightarrow
   \Pi^{KK}_{i_1i_2i_3(\sigma\pmba_1)(\sigma\pmba_2)},
&\nonum\\&
\Pi^{KK}_{i_1i_2\pmba_1\pmba_2\pmba_3}\rightarrow
   \Pi^{KK}_{i_1i_2(\sigma\pmba_1)(\sigma\pmba_2)(\sigma\pmba_3)},\quad
\Pi^{KK}_{i\pmba_1\cdots\pmba_4}\rightarrow
   \Pi^{KK}_{i(\sigma\pmba_1)\cdots(\sigma\pmba_4)},\quad
\Pi^{KK}_{\pmba_1\cdots\pmba_5}\rightarrow
   \Pi^{KK}_{(\sigma\pmba_1)\cdots(\sigma\pmba_5)},
&
\eea
the MC equations (\ref{alg;KK5;1})$\sim$(\ref{alg;KK5;6})
for the KK5-brane charges stay invariant.
We find that the existence of the KK5-brane charges
in the algebra necessarily requires the presence of the charges of
D-strings, F-strings, D3-branes, D5-branes and NS5-branes.
The descending central series are
$\mathfrak h^0=
   sP[ Z^A, \S^A, \S^{ABC}, \S^{A_1\cdots A_5}, Z^{A_1\cdots A_5},
   P^{I_1\cdots I_5} ]$,
$\mathfrak h^1=
   \{ P_a, Z^A, \S^A, \S^{ABC}, \S^{A_1\cdots A_5},
   Z^{A_1\cdots A_5}, P^{I_1\cdots I_5} \}$,
$\mathfrak h^2=
   \{ Z^\pmba, \S^\pmba, \S^{\pmba AB}, \S^{\pmba A_1\cdots A_4},
   Z^{\pmba A_1\cdots A_4}, P^{\pmba I_1\cdots I_4} \}$,
$\mathfrak h^3=
   \{ \S^{\pmba\pmbb A}, \S^{\pmba\pmbb A_1A_2A_3},
   Z^{\pmba\pmbb A_1A_2A_3}, P^{\pmba\pmbb I_1I_2I_3} \}$,
$\mathfrak h^4=
   \{ \S^{\pmba\pmbb\pmbga}, \S^{\pmba\pmbb\pmbga AB},
   Z^{\pmba\pmbb\pmbga AB}, P^{\pmba\pmbb\pmbga IJ} \}$,
$\mathfrak h^5=
   \{ \S^{\pmba_1\cdots\pmba_4 A},
   Z^{\pmba_1\cdots\pmba_4 A}, P^{\pmba_1\cdots\pmba_4 I} \}$,
$\mathfrak h^6=
   \{ \S^{\pmba_1\cdots\pmba_5},
   Z^{\pmba_1\cdots\pmba_5}, P^{\pmba_1\cdots\pmba_5} \}$,
$\mathfrak h^7=
   \{ 0 \}$.
These are the ideals of our algebra, and
$\mathfrak h^6$ is the center of the KK5-brane superalgebra
$sP[Z^A,\S^A,\S^{ABC},\S^{A_1\cdots A_5},Z^{A_1\cdots A_5},
P^{I_1\cdots I_5} ]$.
Considering quotient algebras, one obtains various subalgebras.

In summary, we have obtained the IIB-brane superalgebra.
The S-duality is realized as an automorphism of our algebra.
The generators of the M-algebra are mapped to those of the
IIB-brane superalgebra, except for the D5-brane charge
$\S^{\pmba_1\cdots\pmba_5}$
and the NS5-brane charge $Z^{\pmba_1\cdots\pmba_5}$.

\section{A Description In Twelve-Dimensions}

In the previous section, we obtained the IIB-brane superalgebra
in presence of all of the F-strings, D-strings, D3-branes,
D5-branes, NS5-branes and KK5-branes.
The S-duality is found to be realized as an automorphism of our superalgebra
which was the generalized SO(2) R-symmetry.
In this section we show this SO(2) symmetry is geometrically realized in
twelve-dimensions with signature (11,1).

As the twelve-dimensional $\Gamma$-matrices,
we define the real representation of Spin(11,1)
using ten-dimensional $\gamma$-matrices as:
\bea
\Ga^a=\ga^a\otimes\sigma_3,\qquad
\Ga^\na=\ga^\na\otimes\sigma_3,\qquad
\Ga^\flat=1\otimes\sigma_1.
\eea
The charge conjugation matrix $C$ in twelve-dimensions 
is related to one in ten-dimensions $c$
as $C=c\otimes\sigma_3$.
It follows that
\bea
&&
(\ga^a1)_{\pmba\pmbb}=
   (\Ga^a)_{\pmba\pmbb},\quad
(\ga^a\sigma_3)_{\pmba\pmbb}=
   (\Ga^{\na a})_{\pmba\pmbb},\quad
(\ga^a\sigma_1)_{\pmba\pmbb}=
   (\Ga^{\flat a})_{\pmba\pmbb},\quad
(\ga_{abc}i\sigma_2)_{\pmba\pmbb}=
   (\Ga_{\na\flat abc})_{\pmba\pmbb},
\nonum\\&&
(\ga_{a_1\cdots a_5}1)_{\pmba\pmbb}=
   (\Ga_{a_1\cdots a_5})_{\pmba\pmbb},\quad
(\ga_{a_1\cdots a_5}\sigma_1)_{\pmba\pmbb}=
   (\Ga_{\flat a_1\cdots a_5})_{\pmba\pmbb},\quad
(\ga_{a_1\cdots a_5}\sigma_3)_{\pmba\pmbb}=
   (\Ga_{\na a_1\cdots a_5})_{\pmba\pmbb}.
\eea
In $(11+1)$-dimensions, $p$-branes with $p=1,2 \bmod 4$ can exist
since $(\Ga^{\mu_1\cdots\mu_p})_{\pmba\pmbb}$ is symmetric with respect to
$\pmba$ and $\pmbb$ when $p=1,2 \bmod 4$.
Writing the $p$-form charges in twelve-dimensions as
${\bfZ}_{\mu_1\cdots \mu_p}$,
we find that the IIB-brane charges are expressed in terms of 
${\bfZ}_{\mu_1\cdots \mu_p}$
as
\bea
\ba{lcr@{\,}ll}
\bfZ_A&\Leftrightarrow&
   \hat\Pi^A=& \Pi^A,&\mbox{ Supertranslation}\\
\bfZ^{\na A}&\Leftrightarrow&
   \hat\Pi_{\na A}=& -\Pi_A,& \mbox{ F-string}\\
\bfZ^{\flat A}&\Leftrightarrow&
   \hat\Pi_{\flat A}=& \Pi_A{}',& \mbox{ D-string}\\
\bfZ^{\na\flat ABC}&\Leftrightarrow&
   \hat\Pi_{\na\flat ABC}=& \Pi_{ABC},& \mbox{ D3-brane}\\
\bfZ^{A_1\cdots A_5}&\Leftrightarrow&
   \hat\Pi_{A_1\cdots A_5}=& \Pi^{KK}_{A_1\cdots A_5},& \mbox{ KK5-brane}\\
\bfZ^{\na A_1\cdots A_5}&\Leftrightarrow&
   \hat\Pi_{\na A_1\cdots A_5}=& -\Pi^{NS}_{A_1\cdots A_5},&
\mbox{ NS5-brane}\\
\bfZ^{\flat A_1\cdots A_5}&\Leftrightarrow&
   \hat\Pi_{\flat A_1\cdots A_5}=& \Pi^D_{A_1\cdots A_5},& \mbox{ D5-brane}
\ea
\eea
where the forms with a caret correspond to the dual forms to the charges
of the branes
in twelve-dimensions.
These charge assignments imply that
the IIB-branes are related to the branes in twelve-dimensions
in  such a way that the S-duality is manifestly realized.
The relations are depicted in figure 1.


\begin{figure}
 \bc
  \includegraphics[height=4cm,clip]{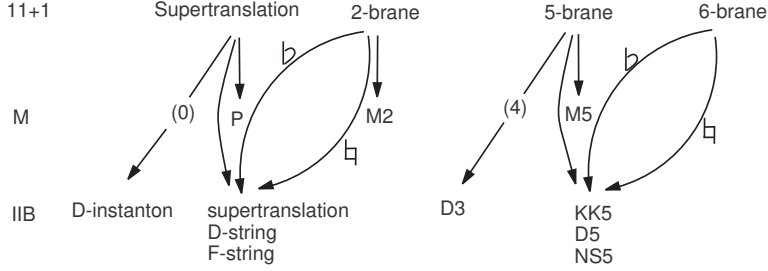}
 \ec
\caption{A unification of the IIB-branes and the M-branes
in twelve-dimensions.}
\end{figure}

In fact, we can rewrite the MC equations of the IIB-branes
as those of the branes in twelve-dimensions.
The MC equations for supertranslations, F-strings and D-strings
are found to be rewritten as (dropping hats)
\bea
d\Pi^a&=&
   -\half\Pi^\pmba\Pi^\pmbb(\Ga^a)_{\pmba\pmbb}
,\\
d\Pi_{ma}&=&
   -\half\Pi^\pmba\Pi^\pmbb(\Ga_{ma})_{\pmba\pmbb}
,\\
d\Pi_{m\pmba}&=&
   -\Pi^\pmbb\Pi_{ma}(\Ga^{a})_{\pmba\pmbb}
   -\Pi^\pmbb\Pi^a(\Ga_{ma})_{\pmba\pmbb}
,
\eea
where $m=\flat, \na$.
The F- and D-strings are unified as the $\na$- and $\flat$-wrapped 2-branes
in twelve-dimensions, respectively.
Projecting the Majorana spinors in twelve-dimensions
to 16+16 Majorana-Weyl spinors with the same chirality
in ten-dimensions,
the above MC equations reduce  to those of supertranslations,
F-strings and D-strings.
The D3-branes
are described as the $(\na,\flat)$-wrapped 5-branes.
The MC equations are found to be rewritten as
\bea
d\Pi_{m_1m_2abc}&=&
   \half\Pi^\pmba\Pi^\pmbb(\Ga_{m_1m_2abc})_{\pmba\pmbb}
,\label{12;3;1}\\
d\Pi_{m_1m_2ab\pmba}&=&
   \Pi^\pmbb\Pi_{m_1m_2abc}(\Ga^c)_{\pmba\pmbb}
   -\Pi^\pmbb\Pi^c(\Ga_{m_1m_2abc})_{\pmba\pmbb}
   +6\Pi^\pmbb\Pi_{m_1m_2}(\Ga_{ab})_{\pmba\pmbb}
,\\
d\Pi_{m_1m_2a\pmba\pmbb}&=&
   -\half\Pi_{m_1m_2abc}\Pi^b(\Ga^c)_{\pmba\pmbb}
   +\frac{1}{4}\Pi_{m_1m_2ab\pmbga}\Pi^\pmbga(\Ga^b)_{\pmba\pmbb}
   +2\Pi_{m_1m_2ab\pmba}\Pi^\pmbga(\Ga^b)_{\pmbb\pmbga}
\nonum\\&&
  -\half\Pi^b\Pi^c(\Ga_{m_1m_2abc})_{\pmba\pmbb}
   +\frac{3}{2}\Pi^b\Pi_{m_1b}(\Ga_{m_2a})_{\pmba\pmbb}
   +3\Pi^b\Pi_{m_1m_2}(\Ga_{ab})_{\pmba\pmbb}
\nonum\\&&
   +\frac{3}{4}\Pi^\pmbga\Pi_{m_1\pmbga}(\Ga_{m_2a})_{\pmba\pmbb}
   -\frac{3}{2}\Pi_{m_1m_2}\Pi_{ab}(\Ga^b)_{\pmba\pmbb}
   +6\Pi^\pmbga\Pi_{m_1\pmba}(\Ga_{m_2a})_{\pmbb\pmbga}
,\\
d\Pi_{m_1m_2\pmba_1\pmba_2\pmba_3}&=&
   2\Pi^a\Pi_{m_1m_2ab\pmba}(\Ga^b)_{\pmba_2\pmba_2}
   +\Pi^\pmbb\Pi_{m_1m_2a\pmbb\pmba_1}(\Ga^a)_{\pmba_2\pmba_3}
   +5\Pi^\pmbb\Pi_{m_1m_2a\pmba_1\pmba_2}(\Ga^a)_{\pmba_3\pmbb}
\nonum\\&&
   -6\Pi_{m_1a}\Pi_{m_2\pmba_1}(\Ga^a)_{\pmba_2\pmba_3}
   +10\Pi^a\Pi_{m_1\pmba_1}(\Ga_{m_2a})_{\pmba_2\pmba_3}
,\label{12;3;4}
\eea
where
it is understood that the symmetries of indices on the left hand side
are to be implemented on the right hand side with unit weight
(anti) symmetrizations.
Projecting the Majorana spinors in twelve-dimensions
to the Majorana-Weyl spinors with the same chirality
in ten-dimensions,
the above MC equations reduce  to those of D3-branes.
The IIB KK5-brane are expected to be described as the 5-branes
in twelve-dimensions.
The MC equations (\ref{alg;KK5;1})$\sim$(\ref{alg;KK5;6})
for KK5-branes 
are rewritten in such a way that the S-duality is geometrically realized.
The MC equation (\ref{alg;KK5;1}) is rewritten as
$
d\Pi_{i_1\cdots i_5}=
   -\half\Pi^\pmba\Pi^\pmbb(\Ga_{i_1\cdots i_5})_{\pmba\pmbb},
$
which is similar to (\ref{12;3;1}).
We expect that
the MC equations for 5-branes in twelve-dimensions
are composed of (\ref{12;3;1})$\sim$(\ref{12;3;4}) and
the MC equations, which is obtained by lifting the MC equations for KK5-branes
into twelve-dimensions.

For NS5- and D5-branes, the corresponding MC equations are found to be
rewritten as
\bea
d\Pi_{ma_1\cdots a_5}&=&
   -\half\Pi^\pmba\Pi^\pmbb(\Ga_{ma_1\cdots a_5})_{\pmba\pmbb}
,\\
d\Pi_{ma_1\cdots a_4\pmba}&=&
   \Pi^\pmbb\Pi_{ma_1\cdots a_4b}(\Ga^b)_{\pmba\pmbb}
   +5\Pi^\pmbb\Pi_{nma_1a_2a_3}(\Ga^n{}_{a_4})_{\pmba\pmbb}
   +\Pi^\pmbb\Pi^b(\Ga_{ma_1\cdots a_4b})_{\pmba\pmbb}
\nonum\\&&
   -5\Pi^\pmbb\Pi_{nm}(\Ga^n{}_{a_1\cdots a_4})_{\pmba\pmbb}
,\\
d\Pi_{ma_1a_2a_3\pmba\pmbb}&=&
   -\half\Pi_{ma_1a_2a_3bc}\Pi^b(\Ga^c)_{\pmba\pmbb}
   +\frac{1}{4}\Pi_{ma_1a_2a_3b\pmbga}\Pi^\pmbga(\Ga^b)_{\pmba\pmbb}
   +2\Pi_{ma_1a_2a_3b\pmba}\Pi^\pmbga(\Ga^b)_{\pmbb\pmbga}
\nonum\\&&
   +2\Pi^b\Pi_{nbma_1a_2}(\Ga^n{}_{a_3})_{\pmba\pmbb}
   +\Pi^b\Pi_{nma_1a_2a_3}(\Ga^n{}_b)_{\pmba\pmbb}
   +2\Pi_{n_1bma_1a_2}\Pi_{n_2a_3}\de^{n_1n_2}(\Ga^b)_{\pmba\pmbb}
\nonum\\&&
   +\half\Pi_{n_1b}\Pi_{n_2ma_1a_2a_3}\de^{n_1n_2}(\Ga^b)_{\pmba\pmbb}
   +\Pi^\pmbga\Pi_{nma_1a_2\pmbga}(\Ga^n{}_{a_3})_{\pmba\pmbb}
   +8\Pi^\pmbga\Pi_{nma_1a_2\pmba}(\Ga^n{}_{a_3})_{\pmbb\pmbga}
\nonum\\&&
   -6\Pi_{nm}\Pi_{a_2a_3}(\Ga^n{}_{a_3})_{\pmba\pmbb}
   -2\Pi^\pmbga\Pi_{n\pmba}(\Ga^n{}_{ma_1a_2a_3})_{\pmbb\pmbga}
   +\half\Pi^b\Pi^c(\Ga_{ma_1a_2a_3bc})_{\pmba\pmbb}
\nonum\\&&
   -6\Pi_{n_1m}\Pi_{n_2a_1}\de^{n_1n_2}(\Ga_{a_2a_3})_{\pmba\pmbb}
   -\half\Pi^b\Pi_{nb}(\Ga^n{}_{ma_1a_2a_3})_{\pmba\pmbb}
   -4\Pi^b\Pi_{nm}(\Ga^n{}_{a_1a_2a_3b})_{\pmba\pmbb}
\nonum\\&&
   -\frac{1}{4}\Pi^\pmbga\Pi_{n\pmbga}(\Ga^n{}_{ma_1a_2a_3})_{\pmba\pmbb}
,\\
d\Pi_{ma_1a_2\pmba_1\pmba_2\pmba_3}&=&
   2\Pi^b\Pi_{ma_1a_2bc\pmba_1}(\Ga^c)_{\pmba_2\pmba_3}
   +\Pi^\pmbb\Pi_{ma_1a_2b\pmbb\pmba_1}(\Ga^b)_{\pmba_2\pmba_3}
   +5\Pi^\pmbb\Pi_{ma_1a_2b\pmba_1\pmba_2}(\Ga^b)_{\pmba_3\pmbb}
\nonum\\&&
   +6\Pi_{n_1m}\Pi_{n_2a_1a_2b\pmba_1}\de^{n_1n_2}(\Ga^b)_{\pmba_2\pmba_3}
   +3\Pi_{n_1b}\Pi_{n_2ma_1a_2\pmba_1}\de^{n_1n_2}(\Ga^b)_{\pmba_2\pmba_3}
\nonum\\&&
   -6\Pi^b\Pi_{nbma_1\pmba_1}(\Ga^n{}_{a_2})_{\pmba_2\pmba_3}
   +5\Pi^b\Pi_{nma_1a_2\pmba_1}(\Ga^n{}_{b})_{\pmba_2\pmba_3}
   +3\Pi^\pmbb\Pi_{nma_1\pmbb\pmba_1}(\Ga^n{}_{a_2})_{\pmba_2\pmba_3}
\nonum\\&&
   +15\Pi^\pmbb\Pi_{nma_1\pmba_1\pmba_2}(\Ga^n{}_{a_2})_{\pmba_3\pmbb}
   +3\Pi_{n_1ma_1a_2b}\Pi_{n_2\pmba_1}\de^{n_1n_2}(\Ga^b)_{\pmba_2\pmba_3}
\nonum\\&&
   +9\Pi_{ma_1}\Pi_{n\pmba_1}(\Ga^n{}_{a_2})_{\pmba_2\pmba_3}
   -12\Pi_{m\pmba_1}\Pi_{na_1}(\Ga^n{}_{a_2})_{\pmba_2\pmba_3}
   +15\Pi_{n_1m}\Pi_{n_2\pmba_1}\de^{n_1n_2}(\Ga_{a_1a_2})_{\pmba_2\pmba_3}
\nonum\\&&
   -5\Pi^b\Pi_{n\pmba_1}(\Ga^n{}_{ma_1a_2b})_{\pmba_2\pmba_3}
,\\
d\Pi_{ma\pmba_1\cdots\pmba_4}&=&
   -\Pi^b\Pi_{mabc\pmba_1\pmba_2}(\Ga^c)_{\pmba_3\pmba_4}
   -\frac{3}{10}\Pi^\pmbb\Pi_{mab\pmbb\pmba_1\pmba_2}(\Ga^b)_{\pmba_3\pmba_4}
   -\frac{6}{5}\Pi^\pmbb\Pi_{mab\pmba_1\pmba_2\pmba_3}(\Ga^b)_{\pmba_4\pmbb}
\nonum\\&&
   +2\Pi_{n_1m}\Pi_{n_2ab\pmba_1\pmba_2}\de^{n_1n_2}(\Ga^b)_{\pmba_3\pmba_4}
   -2\Pi_{n_1b}\Pi_{n_2ma\pmba_1\pmba_2}\de^{n_1n_2}(\Ga^b)_{\pmba_3\pmba_4}
\nonum\\&&
   +2\Pi^b\Pi_{nmb\pmba_1\pmba_2}(\Ga^n{}_{a})_{\pmba_3\pmba_4}
   -3\Pi^b\Pi_{nma\pmba_1\pmba_2}(\Ga^n{}_b)_{\pmba_3\pmba_4}
   -\frac{3}{5}\Pi^\pmbb\Pi_{nm\pmbb\pmba_1\pmba_2}(\Ga^n{}_a)_{\pmba_3\pmba_4}
\nonum\\&&
   -\frac{12}{5}\Pi^\pmbb\Pi_{nm\pmba_1\pmba_2\pmba_3}(\Ga^n{}_a)_{\pmba_4\pmbb}
   +3\Pi_{n_1mab\pmba_1}\Pi_{n_2\pmba_2}\de^{n_1n_2}(\Ga^b)_{\pmba_3\pmba_4}
\nonum\\&&
   -6\Pi_{m\pmba_1}\Pi_{n\pmba_2}(\Ga^n{}_a)_{\pmba_3\pmba_4}
\nonum\\&&
   +3\Pi_{n_1\pmba_1}\Pi_{n_2\pmba_2}\de^{n_1n_2}(\Ga_{ma})_{\pmba_3\pmba_4}
,\\
d\Pi_{m\pmba_1\cdots\pmba_5}&=&
   -\frac{35}{6}\Pi^\pmbb\Pi_{ma\pmba_1\cdots\pmba_4}(\Ga^a)_{\pmba_5\pmbb}
   -\frac{5}{3}\Pi^\pmbb\Pi_{ma\pmbb\pmba_1\pmba_2\pmba_3}(\Ga^a)_{\pmba_4\pmba_5}
   +\Pi^a\Pi_{mab\pmba_1\pmba_2\pmba_3}(\Ga^b)_{\pmba_4\pmba_5}
\nonum\\&&
   +\frac{7}{2}\Pi^a\Pi_{nm\pmba_1\pmba_2\pmba_3}(\Ga^n{}_a)_{\pmba_4\pmba_5}
   +\frac{5}{2}\Pi_{n_1a}\Pi_{n_2m\pmba_1\pmba_2\pmba_3}\de^{n_1n_2}(\Ga^a)_{\pmba_4\pmba_5}
\nonum\\&&
   -\frac{15}{2}\Pi_{n_1\pmba_1}\Pi_{n_2ma\pmba_2\pmba_3}\de^{n_1n_2}(\Ga^a)_{\pmba_4\pmba_5}
.
\eea
The NS5- and D5-branes are unified as the $\flat$- and $\na$-wrapped 6-branes
in twelve-dimensions, respectively.
In this way, we find (non-covariant) twelve-dimensional description
of the IIB-branes. 

The superalgebra is essentially ten-dimensional.
In order to unify the M-algebra and the IIB-brane superalgebra,
one must lift the superalgebra to a superalgebra
with the eleven-dimensional covariance.
The relation of the branes in twelve-dimensions to the M-branes
are described in figure 1.
The eleven-dimensional supertranslations are described in terms of $\bfZ_{M}$.
The 2- and 5-brane charges $\bfZ^{MN}$ and $\bfZ^{M_1\cdots M_5}$
turn out to be the corresponding part of M2- and M5-branes.
One finds that in order to describe the full M-algebra,
we must add the charge $Z_{\pmba\pmbb}$
to the superalgebra obtained
from the IIB superalgebra.
We do not give the unified superalgebra here,
but we expect that IIB-branes and M-branes will
be unified in (11+1)-dimensions.

\section{Summary and Discussions}
We have shown that the WZ terms of D$p$- and NS$p$-branes
in the type-IIB theory
are characterized by non-trivial Chevalley-Eilenberg $(p+2)$-cocycles.
The total actions are equivalent to those obtained in \cite{APS2}
and $\kappa$-invariant.
These WZ terms are shown to be
constructed in a manifestly supersymmetric way
by introducing a set of new spacetime superalgebras.
Using these superalgebras, we have shown that Siegel's
formulation generalizes to the D3-branes, the D5-branes
and the NS5-branes.
Namely, using the supercurrents on the supergroup manifolds
corresponding to the new superalgebras,
we wrote down the WZ terms, which are
$(p+2)$-th order in the supercurrents ($p=5$ for the D3-branes and
$p=7$ for the D5- and NS5-branes).

The relations of these superalgebras
to the M-algebra are discussed.
The D5- and NS5-brane superalgebras contain
generators with five spinorial indices,
which can not be related to the M-algebra.
In order to construct the manifestly superinvariant
WZ terms, these charges must be added.
We have established the IIB-brane superalgebra
with maximal ``central'' extension
which is generated by generators:
supertranslations $Q_A$, D-strings $\S^A$, F-strings $Z^A$,
D3-branes $\S^{ABC}$, D5-branes $\S^{A_1\cdots A_5}$,
NS5-branes $Z^{A_1\cdots A_5}$ and KK5-branes $P^{I_1\cdots I_5}$.
The superalgebra enjoys the S-duality, or generalized SO(2) R-symmetry
as an automorphism of the algebra.
This SO(2)-symmetry is found to be geometrically realized in twelve-dimensions
with signature (11,1).

We expect that our formulation generalizes to
the higher branes: KK5-, D7-, NS7-, D9- and NS9-branes.
The D7- and NS7-brane charges will be contained in $\S^{ABC}$,
and the D9- and NS9-brane charges in $\S^A$ and $Z^A$, respectively.
We have presented the IIB-brane superalgebra
generated by these generators,
but the construction of the corresponding WZ terms
are left for the future investigations.

We have classified IIB NS/D$p$-branes by CE cohomology $(p+2)$-cocycles.
It is interesting for us to consider
whether the CE cohomology classification of KK5-branes
can be achieved.
One wishes to know the physical meaning of the new generators
with spinorial indices.
In \cite{CdAIPB}, the corresponding Noether charges are obtained
and found to probe the topology of the ordinary superspace.
The same will be true for our models too.
In \cite{AHKT},  SO(2,1)$\cong$SL(2,$\mathbb{R}$)
covariant IIB superalgebra was discovered.
There, translations $P_a$, F-strings $Z^a$ and D-strings $\S_a$
formed a SO(2,1) triplet.
It is interesting for us to construct SO(2,1) covariant superalgebra
which contains charges of KK5-branes, NS5-branes and D5-branes
as a SO(2,1) triplet.
If this is achieved, new geometric formulation of $(p,q)$ 5-branes
will be given.

Our formulation will be generalized to the case where the target space
is curved.
For example, one considers the Green-Schwarz superstring actions
on anti-de Sitter spaces: $AdS_3\times S^3$, $AdS_5\times S^5$, etc.
The corresponding coset superspaces are
$\frac{SU(1,1|2)_L\times SU(1,1|2)_R}{SO(2,1)\times SO(3)}$
for the GS strings on $AdS_3\times S^3$
and $\frac{SU(2,2|4)}{SO(4,1)\times SO(5)}$
for the GS strings on $AdS_5\times S^5$.
In order for our formulation to generalize to these cases,
the supergroups $SU(1,1|2)_L\times SU(1,1|2)_R$
or $SU(2,2|4)$ must be extended including new generators.
If this is achieved, the new coordinates corresponding to new generators
will be contained in a surface term.
We hope that the physical meaning of the new coordinates
will be revealed with considering these models,
and our formulation will shed some light on
the various string dualities.

\acknowledgments
The author would like to express his gratitude to Prof. H. Kunitomo
and the members of YITP.
We are also grateful to
Prof. N. Ishibashi and the members of KEK, where this research was started.

\bigskip

\appendix
\section{D5-Brane Superalgebra}\label{D5-brane superalgebra}
The D5-brane superalgebra is generated by
supertranslations $Q_A$, D-strings $\S^A$, F-strings $Z^A$,
D3-branes $\S^{ABC}$ and D5-branes $\S^{A_1\cdots A_5}$.
The MC equations corresponding to the D5-branes are
(\ref{alg;supertranslation}), (\ref{alg;D1}), (\ref{alg;F1}),
(\ref{alg;D3}) and
\bea
d\Pi_{a_1\cdots a_5}^D&=&
   -\half\Pi^\pmba\Pi^\pmbb(\ga_{a_1\cdots a_5}\sigma_1)_{\pmba\pmbb}
,\label{alg;D5-1}\\
d\Pi_{a_1\cdots a_4\pmba}^D&=&
   \Pi^\pmbb\Pi_{a_1\cdots a_4b}^D(\ga^b1)_{\pmba\pmbb}
   +4\Pi^\pmbb\Pi_{a_1a_2a_3}(\ga_{a_4}\sigma_3)_{\pmba\pmbb}
   +\Pi^\pmbb\Pi^b(\ga_{a_1\cdots a_4b}\sigma_1)_{\pmba\pmbb}
\nonum\\&&
   -4\Pi^\pmbb\Pi_{a_1}(\ga_{a_2a_3a_4}i\sigma_2)_{\pmba\pmbb}
,\\
d\Pi_{a_1a_2a_3\pmba\pmbb}^D&=&
   -\half\Pi_{a_1a_2a_3bc}^D\Pi^b(\ga_c1)_{\pmba\pmbb}
   +\frac{1}{4}\Pi_{a_1a_2a_3b\pmbga}^D\Pi^\pmbga(\ga^b1)_{\pmba\pmbb}
   +2\Pi_{a_1a_2a_3b\pmba}^D\Pi^\pmbga(\ga^b1)_{\pmbb\pmbga}
\nonum\\&&
   -\frac{3}{2}\Pi^b\Pi_{ba_1a_2}(\ga_{a_3}\sigma_3)_{\pmba\pmbb}
   +\Pi^b\Pi_{a_1a_2a_3}(\ga_b\sigma_3)_{\pmba\pmbb}
   +\frac{3}{2}\Pi_{ba_1a_2}\Pi_{a_3}(\ga^b1)_{\pmba\pmbb}
\nonum\\&&
   -\half\Pi_b\Pi_{a_1a_2a_3}(\ga^b1)_{\pmba\pmbb}
   +\frac{3}{4}\Pi^\pmbga\Pi_{a_1a_2\pmbga}(\ga_{a_3}\sigma_3)_{\pmba\pmbb}
   +6\Pi^\pmbga\Pi_{a_1a_2\pmba}(\ga_{a_3}\sigma_3)_{\pmbb\pmbga}
\nonum\\&&
   -3\Pi_{a_1}{\Pi_{a_2}}'(\ga_{a_3}\sigma_3)_{\pmba\pmbb}
   +2\Pi^\pmbga\Pi_\pmba(\ga_{a_1a_2a_3}i\sigma_2)_{\pmbb\pmbga}
   +\half\Pi^b\Pi^c(\ga_{a_1a_2a_3bc}\sigma_1)_{\pmba\pmbb}
\nonum\\&&
   -3\Pi_{a_1}\Pi_{a_2}(\ga_{a_3}\sigma_1)_{\pmba\pmbb}
   +\half\Pi^b\Pi_b(\ga_{a_1a_2a_3}i\sigma_2)_{\pmba\pmbb}
   -3\Pi^b\Pi_{a_1}(\ga_{a_2a_3b}i\sigma_2)_{\pmba\pmbb}
\nonum\\&&
   +\frac{1}{4}\Pi^\pmbga\Pi_\pmbga(\ga_{a_1a_2a_3}i\sigma_2)_{\pmba\pmbb}
,\\
d\Pi_{a_1a_2\pmba_1\pmba_2\pmba_3}^D&=&
   2\Pi^b\Pi_{a_1a_2bc\pmba_1}^D(\ga^c1)_{\pmba_2\pmba_3}
   +\Pi^\pmbb\Pi_{a_1a_2b\pmbb\pmba_1}^D(\ga^b1)_{\pmba_2\pmba_3}
   +5\Pi^\pmbb\Pi_{a_1a_2b\pmba_1\pmba_2}^D(\ga^b1)_{\pmba_3\pmbb}
\nonum\\&&
   +4\Pi_{a_1}\Pi_{a_2b\pmba_1}(\ga^b1)_{\pmba_2\pmba_3}
   -3\Pi_b\Pi_{a_1a_2\pmba_1}(\ga^b1)_{\pmba_2\pmba_3}
   +4\Pi^b\Pi_{ba_1\pmba_1}(\ga_{a_2}\sigma_3)_{\pmba_2\pmba_3}
\nonum\\&&
   +5\Pi^b\Pi_{a_1a_2\pmba_1}(\ga_b\sigma_3)_{\pmba_2\pmba_3}
   +2\Pi^\pmbb\Pi_{a_1\pmbb\pmba_1}(\ga_{a_2}\sigma_3)_{\pmba_2\pmba_3}
   +10\Pi^\pmbb\Pi_{a_1\pmba_1\pmba_2}(\ga_{a_2}\sigma_3)_{\pmba_3\pmbb}
\nonum\\&&
   -3\Pi_{a_1a_2b}\Pi_{\pmba_1}(\ga^b1)_{\pmba_2\pmba_3}
   -6{\Pi_{a_1}}'\Pi_{\pmba_1}(\ga_{a_2}\sigma_3)_{\pmba_2\pmba_3}
   +4{\Pi_{\pmba_1}}'\Pi_{a_1}(\ga_{a_2}\sigma_3)_{\pmba_2\pmba_3}
\nonum\\&&
   +5\Pi^b\Pi_{\pmba_1}(\ga_{a_1a_2b}i\sigma_2)_{\pmba_2\pmba_3}
   -10\Pi_{a_1}\Pi_{\pmba_1}(\ga_{a_2}\sigma_1)_{\pmba_2\pmba_3}
,\\
d\Pi_{a\pmba_1\cdots\pmba_4}^D&=&
   -\Pi^b\Pi_{abc\pmba_1\pmba_2}^D(\ga^c1)_{\pmba_3\pmba_4}
   -\frac{3}{10}\Pi^\pmbb\Pi_{ab\pmbb\pmba_1\pmba_2}^D(\ga^b1)_{\pmba_3\pmba_4}
   -\frac{6}{5}\Pi^\pmbb\Pi_{ab\pmba_1\pmba_2\pmba_3}^D(\ga^b1)_{\pmba_4\pmbb}
\nonum\\&&
   +\Pi_a\Pi_{b\pmba_1\pmba_2}(\ga^b1)_{\pmba_3\pmba_4}
   +2\Pi_b\Pi_{a\pmba_1\pmba_2}(\ga^b1)_{\pmba_3\pmba_4}
   +\Pi^b\Pi_{b\pmba_1\pmba_2}(\ga_a\sigma_3)_{\pmba_3\pmba_4}
\nonum\\&&
   -3\Pi^b\Pi_{a\pmba_1\pmba_2}(\ga_b\sigma_3)_{\pmba_3\pmba_4}
   -\frac{3}{10}\Pi^\pmbb\Pi_{\pmbb\pmba_1\pmba_2}
      (\ga_a\sigma_3)_{\pmba_3\pmba_4}
   -\frac{6}{5}\Pi^\pmbb\Pi_{\pmba_1\pmba_2\pmba_3}
      (\ga_a\sigma_3)_{\pmba_4\pmbb}
\nonum\\&&
   -3\Pi_{ab\pmba_1}\Pi_{\pmba_2}(\ga^b1)_{\pmba_3\pmba_4}
   +3{\Pi_{\pmba_1}}'\Pi_{\pmba_2}(\ga_a\sigma_3)_{\pmba_3\pmba_4}
   +3\Pi_{\pmba_1}\Pi_{\pmba_2}(\ga_a\sigma_1)_{\pmba_3\pmba_4}
\label{alg;D5-5}\\
d\Pi_{\pmba_1\cdots\pmba_5}^D&=&
   -\frac{35}{6}\Pi^\pmbb\Pi_{a\pmba_1\cdots\pmba_4}^D(\ga^a1)_{\pmba_5\pmbb}
   -\frac{5}{3}\Pi^\pmbb\Pi_{a\pmbb\pmba_1\pmba_2\pmba_3}^D
      (\ga^a1)_{\pmba_4\pmba_5}
   +\Pi^a\Pi_{ab\pmba_1\pmba_2\pmba_3}^D(\ga^b1)_{\pmba_4\pmba_5}
\nonum\\&&
   +\frac{7}{2}\Pi^a\Pi_{\pmba_1\pmba_2\pmba_3}(\ga_a\sigma_3)_{\pmba_4\pmba_5}
   -\frac{5}{2}\Pi_a\Pi_{\pmba_1\pmba_2\pmba_3}(\ga^a1)_{\pmba_4\pmba_5}
   +\frac{15}{2}\Pi_{\pmba_1}\Pi_{a\pmba_2\pmba_3}(\ga^a1)_{\pmba_4\pmba_5}.
\label{alg;D5-6}
\eea

\section{NS5-Brane Superalgebra}\label{algebra;NS5}
The NS5-brane superalgebra is generated by
supertranslations $Q_A$, D-strings $\S^A$, F-strings $Z^A$,
D3-branes $\S^{ABC}$ and NS5-branes $Z^{A_1\cdots A_5}$.
The MC equations corresponding to the NS5-branes are
(\ref{alg;supertranslation}), (\ref{alg;D1}), (\ref{alg;F1}),
(\ref{alg;D3}) and
\bea
d\Pi_{a_1\cdots a_5}^{NS}&=&
   +\half\Pi^\pmba\Pi^\pmbb(\ga_{a_1\cdots a_5}\sigma_3)_{\pmba\pmbb}
,\label{alg;NS5-1}\\
d\Pi_{a_1\cdots a_4\pmba}^{NS}&=&
   \Pi^\pmbb\Pi_{a_1\cdots a_4b}^{NS}(\ga^b\sigma_3)_{\pmba\pmbb}
   +4\Pi^\pmbb\Pi_{a_1a_2a_3}(\ga_{a_4}\sigma_1)_{\pmba\pmbb}
   -\Pi^\pmbb\Pi^b(\ga_{a_1\cdots a_4b}\sigma_3)_{\pmba\pmbb}
\nonum\\&&
   +4\Pi^\pmbb\Pi_{a_1}{}'(\ga_{a_2a_3a_4}i\sigma_2)_{\pmba\pmbb}
,\\ 
d\Pi_{a_1a_2a_3\pmba\pmbb}^{NS}&=&
   -\half\Pi_{a_1a_2a_3bc}^{NS}\Pi^b(\ga_c1)_{\pmba\pmbb}
   +\frac{1}{4}\Pi_{a_1a_2a_3b\pmbga}^{NS}\Pi^\pmbga(\ga^b1)_{\pmba\pmbb}
   -2\Pi_{a_1a_2a_3b\pmba}^{NS}\Pi^\pmbga(\ga^b1)_{\pmbb\pmbga}
\nonum\\&&
   -\frac{3}{2}\Pi^b\Pi_{ba_1a_2}(\ga_{a_3}\sigma_1)_{\pmba\pmbb}
   +\Pi^b\Pi_{a_1a_2a_3}(\ga_b\sigma_1)_{\pmba\pmbb}
   -\frac{3}{2}\Pi_{ba_1a_2}\Pi_{a_3}{}'(\ga^b1)_{\pmba\pmbb}
\nonum\\&&
   +\half\Pi_b{}'\Pi_{a_1a_2a_3}(\ga^b1)_{\pmba\pmbb}
   +\frac{3}{4}\Pi^\pmbga\Pi_{a_1a_2\pmbga}(\ga_{a_3}\sigma_1)_{\pmba\pmbb}
   +6\Pi^\pmbga\Pi_{a_1a_2\pmba}(\ga_{a_3}\sigma_1)_{\pmbb\pmbga}
\nonum\\&&
   +3\Pi_{a_1}{}'{\Pi_{a_2}}(\ga_{a_3}\sigma_1)_{\pmba\pmbb}
   -2\Pi^\pmbga\Pi_\pmba{}'(\ga_{a_1a_2a_3}i\sigma_2)_{\pmbb\pmbga}
   -\half\Pi^b\Pi^c(\ga_{a_1a_2a_3bc}\sigma_3)_{\pmba\pmbb}
\nonum\\&&
   +3\Pi_{a_1}{}'\Pi_{a_2}{}'(\ga_{a_3}\sigma_3)_{\pmba\pmbb}
   -\half\Pi^b\Pi_b{}'(\ga_{a_1a_2a_3}i\sigma_2)_{\pmba\pmbb}
   +3\Pi^b\Pi_{a_1}{}'(\ga_{a_2a_3b}i\sigma_2)_{\pmba\pmbb}
\nonum\\&&
   -\frac{1}{4}\Pi^\pmbga\Pi_\pmbga{}'(\ga_{a_1a_2a_3}i\sigma_2)_{\pmba\pmbb}
,\\
d\Pi_{a_1a_2\pmba_1\pmba_2\pmba_3}^{NS}&=&
   2\Pi^b\Pi_{a_1a_2bc\pmba_1}^{NS}(\ga^c1)_{\pmba_2\pmba_3}
   +\Pi^\pmbb\Pi_{a_1a_2b\pmbb\pmba_1}^{NS}(\ga^b1)_{\pmba_2\pmba_3}
   +5\Pi^\pmbb\Pi_{a_1a_2b\pmba_1\pmba_2}^{NS}(\ga^b1)_{\pmba_3\pmbb}
\nonum\\&&
   -4\Pi_{a_1}{}'\Pi_{a_2b\pmba_1}(\ga^b1)_{\pmba_2\pmba_3}
   +3\Pi_b{}'\Pi_{a_1a_2\pmba_1}(\ga^b1)_{\pmba_2\pmba_3}
   +4\Pi^b\Pi_{ba_1\pmba_1}(\ga_{a_2}\sigma_1)_{\pmba_2\pmba_3}
\nonum\\&&
   +5\Pi^b\Pi_{a_1a_2\pmba_1}(\ga_b\sigma_1)_{\pmba_2\pmba_3}
   +2\Pi^\pmbb\Pi_{a_1\pmbb\pmba_1}(\ga_{a_2}\sigma_1)_{\pmba_2\pmba_3}
   +10\Pi^\pmbb\Pi_{a_1\pmba_1\pmba_2}(\ga_{a_2}\sigma_1)_{\pmba_3\pmbb}
\nonum\\&&
   +3\Pi_{a_1a_2b}\Pi_{\pmba_1}{}'(\ga^b1)_{\pmba_2\pmba_3}
   +6{\Pi_{a_1}}\Pi_{\pmba_1}{}'(\ga_{a_2}\sigma_1)_{\pmba_2\pmba_3}
   -4{\Pi_{\pmba_1}}\Pi_{a_1}{}'(\ga_{a_2}\sigma_1)_{\pmba_2\pmba_3}
\nonum\\&&
   -5\Pi^b\Pi_{\pmba_1}{}'(\ga_{a_1a_2b}i\sigma_2)_{\pmba_2\pmba_3}
   +10\Pi_{a_1}{}'\Pi_{\pmba_1}{}'(\ga_{a_2}\sigma_3)_{\pmba_2\pmba_3}
,\\
d\Pi_{a\pmba_1\cdots\pmba_4}^{NS}&=&
   -\Pi^b\Pi_{abc\pmba_1\pmba_2}^{NS}(\ga^c1)_{\pmba_3\pmba_4}
   -\frac{3}{10}\Pi^\pmbb\Pi_{ab\pmbb\pmba_1\pmba_2}^{NS}
      (\ga^b1)_{\pmba_3\pmba_4}
   -\frac{6}{5}\Pi^\pmbb\Pi_{ab\pmba_1\pmba_2\pmba_3}^{NS}
      (\ga^b1)_{\pmba_4\pmbb}
\nonum\\&&
   -\Pi_a{}'\Pi_{b\pmba_1\pmba_2}(\ga^b1)_{\pmba_3\pmba_4}
   -2\Pi_b{}'\Pi_{a\pmba_1\pmba_2}(\ga^b1)_{\pmba_3\pmba_4}
   +\Pi^b\Pi_{b\pmba_1\pmba_2}(\ga_a\sigma_1)_{\pmba_3\pmba_4}
\nonum\\&&
   -3\Pi^b\Pi_{a\pmba_1\pmba_2}(\ga_b\sigma_1)_{\pmba_3\pmba_4}
   -\frac{3}{10}\Pi^\pmbb\Pi_{\pmbb\pmba_1\pmba_2}
      (\ga_a\sigma_1)_{\pmba_3\pmba_4}
   -\frac{6}{5}\Pi^\pmbb\Pi_{\pmba_1\pmba_2\pmba_3}
      (\ga_a\sigma_1)_{\pmba_4\pmbb}
\nonum\\&&
   +3\Pi_{ab\pmba_1}\Pi_{\pmba_2}{}'(\ga^b1)_{\pmba_3\pmba_4}
   -3{\Pi_{\pmba_1}}\Pi_{\pmba_2}{}'(\ga_a\sigma_1)_{\pmba_3\pmba_4}
   -3\Pi_{\pmba_1}{}'\Pi_{\pmba_2}{}'(\ga_a\sigma_3)_{\pmba_3\pmba_4}
,\label{alg;NS5-5}\\
d\Pi_{\pmba_1\cdots\pmba_5}^{NS}&=&
   -\frac{35}{6}\Pi^\pmbb\Pi_{a\pmba_1\cdots\pmba_4}^{NS}
      (\ga^a1)_{\pmba_5\pmbb}
   -\frac{5}{3}\Pi^\pmbb\Pi_{a\pmbb\pmba_1\pmba_2\pmba_3}^{NS}
      (\ga^a1)_{\pmba_4\pmba_5}
   +\Pi^a\Pi_{ab\pmba_1\pmba_2\pmba_3}^{NS}(\ga^b1)_{\pmba_4\pmba_5}
\nonum\\&&
   +\frac{7}{2}\Pi^a\Pi_{\pmba_1\pmba_2\pmba_3}
     (\ga_a\sigma_1)_{\pmba_4\pmba_5}
   +\frac{5}{2}\Pi_a{}'\Pi_{\pmba_1\pmba_2\pmba_3}
     (\ga^a1)_{\pmba_4\pmba_5}
   -\frac{15}{2}\Pi_{\pmba_1}{}'\Pi_{a\pmba_2\pmba_3}
     (\ga^a1)_{\pmba_4\pmba_5}
.\label{alg;NS5-6}
\eea

\section{KK5-Brane Superalgebra}\label{algebra;KK5}
The KK5-brane superalgebra is generated by
supertranslations $Q_A$, D-strings $\S^A$, F-strings $Z^A$,
D3-branes $\S^{ABC}$, D5-branes $\S^{A_1\cdots A_5}$,
NS5-branes $Z^{A_1\cdots A_5}$ and KK5-branes $P^{A_1\cdots A_5}$.
The MC equations needed for the KK5-branes are
(\ref{alg;supertranslation}), (\ref{alg;D1}), (\ref{alg;F1}),
(\ref{alg;D3}), (\ref{alg;D5-1}-\ref{alg;D5-6}),
(\ref{alg;NS5-1}-\ref{alg;NS5-6}) and
\bea
d\Pi_{i_1\cdots i_5}^{KK}&=&
   -\half\Pi^\pmba\Pi^\pmbb(\ga_{i_1\cdots i_5}1)_{\pmba\pmbb},
\label{alg;KK5;1}\\
d\Pi_{i_1\cdots i_4\pmba}^{KK}&=&
   \Pi^\pmbb\Pi^j(\ga_{ji_1\cdots i_4}1)_{\pmba\pmbb}
   -\Pi^\pmbb\Pi_\sh(\ga_{\sh i_1\cdots i_4}\sigma_3)_{\pmba\pmbb}
   +\Pi^\pmbb\Pi_\sh{}'(\ga_{i_1\cdots i_4\sh}\sigma_1)_{\pmba\pmbb}
\nonum\\&&
   +\Pi^\pmbb\Pi_{ji_1\cdots i_4}^{KK}(\ga^j1)_{\pmba\pmbb}
   -\Pi^\pmbb\Pi_{\sh i_1\cdots i_4}^{NS}(\ga^\sh\sigma_3)_{\pmba\pmbb}
   +\Pi^\pmbb\Pi_{\sh i_1\cdots i_4}^{D}(\ga^\sh\sigma_1)_{\pmba\pmbb}
\nonum\\&&
   +6\Pi^\pmbb\Pi_{\sh i_1i_2}(\ga_{\sh i_3i_4}i\sigma_2)_{\pmba\pmbb},
\\
d\Pi_{i_1i_2i_3\pmba\pmbb}^{KK}&=&
   \half\Pi^j\Pi^k(\ga_{jki_1i_2i_3}1)_{\pmba\pmbb}
   -\Pi_\sh\Pi^j(\ga_{ji_1i_2i_3\sh}\sigma_3)_{\pmba\pmbb}
   +\Pi_\sh{}'\Pi^j(\ga_{ji_1i_2i_3\sh}\sigma_1)_{\pmba\pmbb}
\nonum\\&&
   -\half\Pi_{i_1i_2i_3jk}^{KK}\Pi^j(\ga^k1)_{\pmba\pmbb}
   +\half\Pi_{\sh i_1i_2i_3j}^{NS}\Pi_\sh(\ga^j1)_{\pmba\pmbb}
   +\half\Pi^D_{\sh i_1i_2i_3j}\Pi_\sh{}'(\ga^j1)_{\pmba\pmbb}
\nonum\\&&
   +\half\Pi^{NS}_{\sh i_1i_2i_3j}\Pi^j(\ga^\sh\sigma_3)_{\pmba\pmbb}
   -\half\Pi^D_{\sh i_1i_2i_3j}\Pi^j(\ga^\sh\sigma_1)_{\pmba\pmbb}
   -\half\Pi_{i_1i_2i_3}\Pi_\sh{}'(\ga^\sh\sigma_3)_{\pmba\pmbb}
\nonum\\&&
   -\half\Pi_{i_1i_2i_3}\Pi_\sh(\ga^\sh\sigma_1)_{\pmba\pmbb}
   +\frac{1}{4}\Pi^{KK}_{i_1i_2i_3j\pmbga}\Pi^\pmbga(\ga^j1)_{\pmba\pmbb}
   +\frac{1}{4}\Pi^{NS}_{\sh i_1i_2i_3\pmbga}
      \Pi^\pmbga(\ga^\sh\sigma_3)_{\pmba\pmbb}
\nonum\\&&
   -\frac{1}{4}\Pi^{D}_{\sh i_1i_2i_3\pmbga}
      \Pi^\pmbga(\ga^\sh\sigma_1)_{\pmba\pmbb}
   -2\Pi^{KK}_{ji_1i_2i_3\pmba}\Pi^\pmbga(\ga^j1)_{\pmbb\pmbga}
   +2\Pi^{NS}_{\sh i_1i_2i_3\pmba}\Pi^\pmbga(\ga^\sh\sigma_3)_{\pmbb\pmbga}
\nonum\\&&
   -2\Pi^D_{\sh i_1i_2i_3\pmba}\Pi^\pmbga(\ga^\sh\sigma_1)_{\pmbb\pmbga}
   -3\Pi_\sh{}'\Pi_{\sh i_1i_2}(\ga_{i_3}\sigma_3)_{\pmba\pmbb}
   -3\Pi_\sh\Pi_{\sh i_1i_2}(\ga_{i_3}\sigma_1)_{\pmba\pmbb}
\nonum\\&&
   -6\Pi^\pmbga\Pi_{\sh i_1\pmba}(\ga_{\sh i_2i_3}i\sigma_2)_{\pmbb\pmbga}
   +\frac{3}{2}\Pi_{\sh ji_1}\Pi_{\sh i_2i_3}(\ga^j1)_{\pmba\pmbb}
   +\frac{3}{2}\Pi_{i_1}{}'\Pi_{\sh i_2i_3}(\ga^\sh\sigma_3)_{\pmba\pmbb}
\nonum\\&&
   +\frac{3}{2}\Pi_{i_1}\Pi_{\sh i_2i_3}(\ga^\sh\sigma_1)_{\pmba\pmbb}
   +\Pi_\sh{}'\Pi_\sh(\ga_{i_1i_2i_3}i\sigma_2)_{\pmba\pmbb}
   -\frac{3}{2}\Pi^j\Pi_{\sh ji_1}(\ga_{i_2i_3\sh}i\sigma_2)_{\pmba\pmbb}
\nonum\\&&
   +\frac{3}{2}\Pi_\sh\Pi_{i_1}{}'(\ga_{i_2i_3\sh}i\sigma_2)_{\pmba\pmbb}
   -\frac{3}{2}\Pi_\sh{}'\Pi_{i_1}(\ga_{i_2i_3\sh}i\sigma_2)_{\pmba\pmbb}
   +3\Pi^j\Pi_{\sh i_1i_2}(\ga_{i_3j\sh}i\sigma_2)_{\pmba\pmbb}
\nonum\\&&
   -\frac{3}{4}\Pi^\ga\Pi_{\sh i_1\pmbga}
      (\ga_{i_2i_3\sh}i\sigma_2)_{\pmba\pmbb},
\\
d\Pi^{KK}_{i_1i_2\pmba_1\pmba_2\pmba_3}&=&
   2\Pi^i\Pi^{KK}_{i_1i_2jk\pmba_1}(\ga^k1)_{\pmba_2\pmba_3}
   +2\Pi_\sh\Pi^{NS}_{\sh i_1i_2j\pmba_1}(\ga^j1)_{\pmba_2\pmba_3}
   +2\Pi_\sh{}'\Pi^{D}_{\sh i_1i_2j\pmba_1}(\ga^j1)_{\pmba_2\pmba_3}
\nonum\\&&
   +2\Pi^j\Pi^{NS}_{\sh i_1i_2j\pmba_1}(\ga^\sh\sigma_3)_{\pmba_2\pmba_3}
   -2\Pi^j\Pi^{D}_{\sh i_1i_2j\pmba_1}(\ga^\sh\sigma_1)_{\pmba_2\pmba_3}
   +2\Pi_\sh{}'\Pi_{i_1i_2\pmba_1}(\ga^\sh\sigma_3)_{\pmba_2\pmba_3}
\nonum\\&&
   +2\Pi_\sh\Pi_{i_1i_2\pmba_1}(\ga^\sh\sigma_1)_{\pmba_2\pmba_3}
   +\Pi^\pmbb\Pi^{KK}_{i_1i_2j\pmbb\pmba_1}(\ga^j1)_{\pmba_2\pmba_3}
   -\Pi^\pmbb\Pi^{NS}_{\sh i_1i_2\pmbb\pmba_1}
      (\ga^\sh\sigma_3)_{\pmba_2\pmba_3}
\nonum\\&&
   +\Pi^\pmbb\Pi^{D}_{\sh i_1i_2\pmbb\pmba_1}
      (\ga^\sh\sigma_1)_{\pmba_2\pmba_3}
   +5\Pi^\pmbb\Pi^{KK}_{i_1i_2j\pmba_1\pmba_2}(\ga^j1)_{\pmba_3\pmbb}
   -5\Pi^\pmbb\Pi^{NS}_{\sh i_1i_2\pmba_1\pmba_2}
      (\ga^\sh\sigma_3)_{\pmba_3\pmbb}
\nonum\\&&
   +5\Pi^\pmbb\Pi^{D}_{\sh i_1i_2\pmba_1\pmba_2}
      (\ga^\sh\sigma_1)_{\pmba_3\pmbb}
   +10\Pi_\sh{}'\Pi_{\sh i_1\pmba_1}(\ga_{i_2}\sigma_3)_{\pmba_2\pmba_3}
   +10\Pi_\sh\Pi_{\sh i_1\pmba_1}(\ga_{i_2}\sigma_1)_{\pmba_2\pmba_3}
\nonum\\&&
   +5\Pi^\pmbb\Pi_{\sh\pmba_1\pmba_2}
     (\ga_{\sh i_1i_2}i\sigma_2)_{\pmba_3\pmbb}
   +6\Pi_{\sh ji_1}\Pi_{\sh i_2\pmba_1}(\ga^j1)_{\pmba_2\pmba_3}
   +6\Pi_{i_1}{}'\Pi_{\sh i_2\pmba_1}(\ga^\sh\sigma_3)_{\pmba_2\pmba_3}
\nonum\\&&
   +6\Pi_{i_1}\Pi_{\sh i_2\pmba_1}(\ga^\sh\sigma_1)_{\pmba_2\pmba_3}
   +2\Pi_{\sh j\pmba_1}\Pi_{\sh i_1i_2}(\ga^j1)_{\pmba_2\pmba_3}
   -2\Pi_{\pmba_1}{}'\Pi_{\sh i_1i_2}(\ga^\sh\sigma_3)_{\pmba_2\pmba_3}
\nonum\\&&
   -2\Pi_{\pmba_1}\Pi_{\sh i_1i_2}(\ga^\sh\sigma_1)_{\pmba_2\pmba_3}
   -10\Pi^j\Pi_{\sh i_1\pmba_1}(\ga_{i_2j\sh}i\sigma_2)_{\pmba_2\pmba_3}
   -2\Pi^j\Pi_{\sh j\pmba_1}(\ga_{\sh i_1i_2}i\sigma_2)_{\pmba_2\pmba_3}
\nonum\\&&
   -2\Pi_\sh\Pi_{\pmba_1}{}'(\ga_{\sh i_1i_2}i\sigma_2)_{\pmba_2\pmba_3}
   +2\Pi_\sh{}'\Pi_{\pmba_1}(\ga_{\sh i_1i_2}i\sigma_2)_{\pmba_2\pmba_3}
   +\Pi^\pmbb\Pi_{\sh\pmbb\pmba_1}
     (\ga_{\sh i_1i_2}i\sigma_2)_{\pmba_2\pmba_3},
\\
d\Pi^{KK}_{i\pmba_1\cdots\pmba_4}&=&
   -\Pi^j\Pi^{KK}_{ijk\pmba_1\pmba_2}(\ga^k1)_{\pmba_3\pmba_4}
   +\Pi^j\Pi^{NS}_{\sh ij\pmba_1\pmba_2}(\ga^\sh\sigma_3)_{\pmba_3\pmba_4}
   -\Pi^j\Pi^{D}_{\sh ij\pmba_1\pmba_2}(\ga^\sh\sigma_1)_{\pmba_3\pmba_4}
\nonum\\&&
   +\Pi_\sh\Pi^{NS}_{\sh ij\pmba_1\pmba_2}(\ga^j1)_{\pmba_3\pmba_4}
   +\Pi_\sh{}'\Pi^{D}_{\sh ij\pmba_1\pmba_2}(\ga^j1)_{\pmba_3\pmba_4}
   -\Pi_\sh{}'\Pi_{i\pmba_1\pmba_2}(\ga^\sh\sigma_3)_{\pmba_3\pmba_4}
\nonum\\&&
   -\Pi_\sh\Pi_{i\pmba_1\pmba_2}(\ga^\sh\sigma_1)_{\pmba_3\pmba_4}
   -\frac{3}{10}\Pi^\pmbb\Pi^{KK}_{ij\pmbb\pmba_1\pmba_2}
       (\ga^j1)_{\pmba_3\pmba_4}
   -\frac{3}{10}\Pi^\pmbb\Pi^{NS}_{\sh i\pmbb\pmba_1\pmba_2}
       (\ga^\sh\sigma_3)_{\pmba_3\pmba_4}
\nonum\\&&
   +\frac{3}{10}\Pi^\pmbb\Pi^{D}_{\sh i\pmbb\pmba_1\pmba_2}
       (\ga^\sh\sigma_1)_{\pmba_3\pmba_4}
   -\frac{6}{5}\Pi^\pmbb\Pi^{KK}_{ij\pmba_1\pmba_2\pmba_3}
       (\ga^j1)_{\pmba_4\pmbb}
   -\frac{6}{5}\Pi^\pmbb\Pi^{NS}_{\sh i\pmba_1\pmba_2\pmba_3}
       (\ga^\sh\sigma_3)_{\pmba_4\pmbb}
\nonum\\&&
   +\frac{6}{5}\Pi^\pmbb\Pi^{D}_{\sh i\pmba_1\pmba_2\pmba_3}
       (\ga^\sh\sigma_1)_{\pmba_4\pmbb}
   +3\Pi_\sh{}'\Pi_{\sh\pmba_1\pmba_2}(\ga_i\sigma_3)_{\pmba_3\pmba_4}
   +3\Pi_\sh\Pi_{\sh\pmba_1\pmba_2}
       (\ga_i\sigma_1)_{\pmba_3\pmba_4}
\nonum\\&&
   +2\Pi_{\sh ij}\Pi_{\sh\pmba_1\pmba_2}(\ga^j1)_{\pmba_3\pmba_4}
   -2\Pi_i{}'\Pi_{\sh\pmba_1\pmba_2}(\ga^\sh\sigma_3)_{\pmba_3\pmba_4}
   -2\Pi_i\Pi_{\sh\pmba_1\pmba_2}(\ga^\sh\sigma_1)_{\pmba_3\pmba_4}
\nonum\\&&
   +3\Pi_{\sh j\pmba_1}\Pi_{\sh i\pmba_2}(\ga^j1)_{\pmba_3\pmba_4}
   -3\Pi_{\pmba_1}{}'\Pi_{\sh i\pmba_2}(\ga^\sh\sigma_3)_{\pmba_3\pmba_4}
   -3\Pi_{\pmba_1}\Pi_{\sh i\pmba_2}(\ga^\sh\sigma_1)_{\pmba_3\pmba_4}
\nonum\\&&
   -3\Pi^j\Pi_{\sh\pmba_1\pmba_2}(\ga_{\sh ij}i\sigma_2)_{\pmba_3\pmba_4},
\\
d\Pi^{KK}_{\pmba_1\cdots\pmba_5}&=&
   -\frac{3}{5}\Pi^i\Pi^{KK}_{ij\pmba_1\pmba_2\pmba_3}
     (\ga^j1)_{\pmba_4\pmba_5}
   -\frac{3}{5}\Pi^i\Pi^{NS}_{\sh i\pmba_1\pmba_2\pmba_3}
     (\ga^\sh\sigma_3)_{\pmba_4\pmba_5}
   +\frac{3}{5}\Pi^i\Pi^{D}_{\sh i\pmba_1\pmba_2\pmba_3}
     (\ga^\sh\sigma_1)_{\pmba_4\pmba_5}
\nonum\\&&
   -\frac{3}{5}\Pi_\sh\Pi^{NS}_{\sh i\pmba_1\pmba_2\pmba_3}
      (\ga^i1)_{\pmba_4\pmba_5}
   -\frac{3}{5}\Pi_\sh{}'\Pi^{D}_{\sh i\pmba_1\pmba_2\pmba_3}
      (\ga^i1)_{\pmba_4\pmba_5}
   -\frac{3}{5}\Pi_\sh{}'\Pi_{\pmba_1\pmba_2\pmba_3}
      (\ga^\sh\sigma_3)_{\pmba_4\pmba_5}
\nonum\\&&
   -\frac{3}{5}\Pi_\sh\Pi_{\pmba_1\pmba_2\pmba_3}
      (\ga^\sh\sigma_1)_{\pmba_4\pmba_5}
   +\Pi^\pmbb\Pi^{KK}_{i\pmbb\pmba_1\pmba_2\pmba_3}
      (\ga^i1)_{\pmba_4\pmba_5}
   -\Pi^\pmbb\Pi^{NS}_{\sh\pmbb\pmba_1\pmba_2\pmba_3}
      (\ga^\sh\sigma_3)_{\pmba_4\pmba_5}
\nonum\\&&
   +\Pi^\pmbb\Pi^{D}_{\sh\pmbb\pmba_1\pmba_2\pmba_3}
      (\ga^\sh\sigma_1)_{\pmba_4\pmba_5}
   +\frac{7}{2}\Pi^\pmbb\Pi^{KK}_{i\pmba_1\cdots\pmba_4}
      (\ga^i1)_{\pmba_5\pmbb}
   -\frac{7}{2}\Pi^\pmbb\Pi^{NS}_{\sh\pmba_1\cdots\pmba_4}
      (\ga^\sh\sigma_3)_{\pmba_5\pmbb}
\nonum\\&&
   +\frac{7}{2}\Pi^\pmbb\Pi^{D}_{\sh\pmba_1\cdots\pmba_4}
      (\ga^\sh\sigma_1)_{\pmba_5\pmbb}
   -6\Pi_{\sh i\pmba_1}\Pi_{\sh\pmba_2\pmba_3}
      (\ga^i1)_{\pmba_4\pmba_5}
   +6\Pi_{\pmba_1}{}'\Pi_{\sh\pmba_2\pmba_3}
       (\ga^\sh\sigma_3)_{\pmba_4\pmba_5}
\nonum\\&&
   +6\Pi_{\pmba_1}\Pi_{\sh\pmba_2\pmba_3}
       (\ga^\sh\sigma_1)_{\pmba_4\pmba_5}.
\label{alg;KK5;6}
\eea


\end{document}